# Correlation Tensor Magnetic Resonance Imaging

Rafael Neto Henriques[1], Sune N Jespersen[2,3] and Noam Shemesh[1*]

[1]Champalimaud Research, Champalimaud Centre for the Unknown, Lisbon, Portugal

[2]Center of Functionally Integrative Neuroscience (CFIN) and MINDLab, Department of Clinical Medicine, Aarhus University, Aarhus, Denmark.

[3]Department of Physics and Astronomy, Aarhus University, Aarhus, Denmark

*Corresponding author:

Dr. Noam Shemesh, Champalimaud Research, Champalimaud Centre for the Unknown, Av. Brasilia 1400-038, Lisbon, Portugal

E-mail: noam.shemesh@neuro.fchampalimaud.org;

Phone number: +351 210 480 000 ext. #4467

# Abstract

Diffusional Kurtosis Magnetic Resonance Imaging (DKI) quantifies the extent of non-Gaussian water diffusion, which has been shown to be a sensitive biomarker for microstructure in health and disease. However, DKI is not specific to any microstructural property per se since kurtosis may emerge from several different sources. Q-space trajectory encoding schemes have been proposed for decoupling kurtosis arising from the variance of mean diffusivities (isotropic kurtosis) from kurtosis driven by microscopic anisotropy (anisotropic kurtosis). Still, these methods assume that the system is comprised of multiple Gaussian diffusion components with vanishing intra-compartmental kurtosis (associated with restricted diffusion). Here, we develop a more general framework for resolving the underlying kurtosis sources without relying on the multiple Gaussian diffusion approximation. We introduce Correlation Tensor MRI (CTI) – an approach harnessing the versatility of double diffusion encoding (DDE) and its sensitivity to displacement correlation tensors capable of explicitly decoupling isotropic and anisotropic kurtosis components from intra-compartmental kurtosis effects arising from restricted (and time-dependent) diffusion. Additionally, we show that, by subtracting these isotropic and anisotropic kurtosis components from the total diffusional kurtosis, CTI provides an index that is potentially sensitive to intra-compartmental kurtosis. The theoretical foundations of CTI, as well as the first proof-of-concept CTI *ex vivo* experiments in mouse brains at ultrahigh field of 16.4 T, are presented. We find that anisotropic and isotropic kurtosis can decouple microscopic anisotropy from substantial partial volume effects between tissue and free water. Our intra-compartmental kurtosis index exhibited positive values in both white and grey matter tissues. Simulations in different synthetic microenvironments show, however, that our current CTI protocol for estimating intra-compartmental kurtosis is limited by higher order terms that were not taken into account in this study. CTI measurements were then extended to *in vivo* settings and used to map heathy rat brains at 9.4 T. These *in vivo* CTI results were found to be consistent with our *ex vivo* findings. Although future studies are still required to assess and mitigate the higher order effects on the intra-compartmental kurtosis index, our results show that CTI's more general estimates of anisotropic and isotropic kurtosis contributions are already ripe for future *in vivo* studies, which can have significant impact our understanding of the mechanisms underlying diffusion metrics extracted in health and disease.

# 1. Introduction

Sensing microstructural features of biological systems noninvasively is vital for understanding how large-scale biological systems evolve over time. Tissue microarchitecture is constantly remodelled, whether due to normal processes such as development, learning and aging (Falangola et al., 2008; Moseley, 2002; Neil et al., 1998; Pfefferbaum et al., 2000), or due to abnormal processes such as disease progression or acute insults to the tissue (Cheung et al., 2012; Fieremans et al., 2013; Moseley et al., 1990a). Often, the microstructural changes precede functional outcomes: for example, spine density increases rapidly before learning has taken place (Xu et al., 2009); subtle changes in cellular density and structure precede the functional deficits incurred in neurological disorders (Hanisch and Kettenmann, 2007); and malignant transformations can occur well before tumours can be detected (Peinado et al., 2017). All these reinforce the need for accurate *in vivo* mapping of microstructural properties (Le Bihan and Johansen-Berg, 2012).

Diffusion MRI (dMRI (Le Bihan and Johansen-Berg, 2012)), mainly based on variants of the Single Diffusion Encoding (SDE, Shemesh et al., 2016) methodology developed originally by Stejskal and Tanner (Stejskal and Tanner, 1965), has become the mainstay of contemporary non-invasive microstructural imaging. Water molecules traverse microscopic length scales on typical MR-relevant observation times at body temperature, and their diffusion properties are influenced by the presence of restricting boundaries, such as cell membranes and other subcellular structures. dMRI capitalizes on this "endogenous sensor" by sensitizing the MRI signal towards molecular displacements in a given orientation, thereby enabling the quantification of water diffusion properties (Assaf and Cohen, 1998; Jensen et al., 2005; Moseley et al., 1990a). In many cases, it is assumed that water diffusion can be fully characterized by a single apparent diffusion tensor (Basser et al., 1994). Diffusion Tensor

Imaging (DTI) can extract this apparent tensor from multiple diffusion-weighted measurements; the tensor's trace has been shown to be sensitive, for example, towards early phases of acute stroke (Reith et al., 1995), while the tensor's orientation can be used to recover the absolute orientation of coherently aligned white matter tracts (Catani et al., 2002; Jones, 2008; Mori et al., 1999), which has been instrumental to, e.g., surgical planning (Berman, 2009).

DTI and similar methods represent the dMRI signal as being sufficiently well characterized by Gaussian diffusion (Basser, 1995; Dell'Acqua et al., 2007; Descoteaux et al., 2009; Tournier et al., 2007). Implicitly, this means that the (logarithm of the) diffusion signal is represented only up to first order in b-value, where the b-value represents the strength of diffusion weighting (Le Bihan et al., 1986; Le Bihan and Breton, 1985). However, it has been recognized already early on that diffusion in biological systems is generally non-Gaussian (Assaf and Cohen, 1998; Mulkern et al., 1999; Sukstanskii and Yablonskiy, 2002; Yablonskiy et al., 2003), and that characterizing the non-Gaussian effects may provide deeper insights into tissue microstructure. To capture non-Gaussian diffusion effects via signal representations, Jensen et al. developed diffusional kurtosis imaging (DKI) (Jensen et al., 2005). In DKI, the signal is expanded using cumulants up to second order in b-value, quantifying the leading deviation from Gaussian diffusion and giving rise to a source of contrast based on non-Gaussian properties of the signal. DKI has been shown to be more sensitive than its DTI counterpart towards quantifying microstructural changes related to, e.g. aging, development and disease (Cheung et al., 2012; Falangola et al., 2008; Fieremans et al., 2013; Gong et al., 2013; Helpern et al., 2011; Hui et al., 2012; Rudrapatna et al., 2014; Wang et al., 2011).

Despite the utility of DTI and DKI, both methods conflate mesoscopic orientation dispersion and true microstructural properties (De Santis et al., 2014; Henriques et al., 2015; Jones et al., 2013; Szczepankiewicz et al., 2015). For example, consider a system with an ensemble of microscopic components (or microenvironments) represented by diffusion tensors with identical trace and anisotropy, dispersed along a mesoscopic orientation distribution (Fig. 1A). The microscopic features in this system consist of microscopic anisotropy (the anisotropy of the microenvironment tensors in their own eigenframe), the distribution of diffusion tensor traces (in the system considered above, just a delta function), and the extent of restricted diffusion (in the system above – none). The main mesoscopic feature is the orientation dispersion (the level of coherence between the different orientations of the otherwise identical tensors). In such a system, the total diffusion anisotropy as measured by DTI is highly dependent on the degree of the mesoscopic orientation dispersion and can span the full range between zero (completely randomly oriented tensors) and one (no orientation dispersion). For any finite orientation dispersion, anisotropy will in general appear smaller in DTI compared with the anisotropy of the underlying microscopic components. The diffusion signal decay in this system will also exhibit non-Gaussian characteristics at higher b-values, with kurtosis emerging from both the microscopic properties and the orientation dispersion. However, no link between kurtosis values and microscopic features can be established without imposing priors about the underlying tissue (Fieremans et al., 2011; Henriques et al., 2019; Jensen et al., 2005; Jensen and Helpern, 2010).

More generally, kurtosis derived from SDE methods is inherently limited in its specificity towards microstructure since it can emerge from multiple sources such as:

(i) anisotropic diffusion within orientationally dispersed yet otherwise identical microenvironments (e.g., an ensemble of identical diffusion tensors dispersed along an orientation distribution) (Henriques et al., 2019; Kaden et al., 2016; Kroenke et al., 2004; Szczepankiewicz et al., 2015; Yablonskiy and Sukstanskii, 2010) – Fig. 1A;

(ii) diffusion arising from a distribution of Gaussian components (isotropic tensors) in different microenvironments (e.g., an ensemble of isotropic tensors that are polydisperse in tensor magnitude) (Fieremans et al., 2011; Jensen et al., 2005; Jensen and Helpern, 2010; Sukstanskii and Yablonskiy, 2002) – Fig. 1B;

(iii) restricted, time-dependent non-Gaussian diffusion (e.g., diffusion within reflecting barriers) (Callaghan, 1995; Callaghan et al., 1991; Dhital et al., 2018; Henriques et al., 2019; Jespersen, 2018; Paulsen et al., 2015) – Fig. 1C.

Kurtosis from all the above sources can be modulated by other factors such as exchange (Jensen et al., 2005; Jensen and Helpern, 2010; Kärger, 1985; Ning et al., 2018) or e.g. surface relaxation. In realistic tissues diffusional kurtosis can arise from a combination of different sources (Fig. 1D), however, current state-of-the-art SDE methods cannot resolve their differential contributions (De Santis et al., 2014; Henriques et al., 2019, 2015; Jones et al., 2013).

Attempting to increase the specificity of diffusion-driven information, several microstructural models have been proposed for directly relating diffusion-weighted signals to tissue properties (Jelescu and Budde, 2017; Nilsson et al., 2013; Novikov et al., 2018a; Yablonskiy and Sukstanskii, 2010). However, due to flat fitting landscapes, several assumptions and constraints are necessary for stabilizing model fitting (Assaf et al., 2004; Fieremans et al., 2011; Jelescu et al., 2016; Jespersen et al., 2007; Stanisz et al., 1997; Zhang et al., 2012). Recent studies have shown that the underlying assumptions and constraints imposed by these methods can significantly compromise the specificity of the extracted parameters (Henriques et al., 2019; Lampinen et al., 2019, 2017; Novikov et al., 2018b).

As an alternative to microstructural models, more specific diffusion characterization can be obtained using advanced pulse sequences (Mitra, 1995; Shemesh et al., 2016; Wong et al., 1995). Double diffusion encoding (DDE) sequences have been proposed to measure microscopic anisotropy independently of mesoscopic orientation dispersion (Callaghan and Komlosh, 2002; Cory et al., 1990; Hui and Jensen, 2015; Jespersen et al., 2013; Lawrenz and Finsterbusch, 2013; Mitra, 1995; Shemesh and Cohen, 2011). Multi-dimensional diffusion encoding (MDE) sequences (de Almeida Martins and Topgaard, 2016; Eriksson et al., 2015, 2013; Topgaard, 2015; Valette et al., 2012; Wong et al., 1995) were also developed for similar purposes, with the explicit assumption that diffusion in tissues can be represented by a sum of non-exchanging Gaussian diffusion components. In particular, q-space trajectory encoding (QTE) was proposed to resolve kurtosis into two different sources (Lasič et al., 2014; Szczepankiewicz et al., 2019, 2016, 2015; Topgaard, 2019, 2017; Westin et al., 2016): 1) anisotropic kurtosis $K_{aniso}$ which is related with the non-Gaussian signal decay arising from microscopic anisotropy; and 2) isotropic kurtosis $K_{iso}$ which underpins the non-Gaussian signal decay arising from the variance of mean diffusivities (polydispersity in diffusion tensor magnitudes). Although recent studies showed that disentangling these two sources of kurtosis can be useful to contrast tissues with disparate microstructural features such as different tumour types (Nilsson et al., 2020; Szczepankiewicz et al., 2016, 2015) the validity of the extracted measures may be compromised by factors not considered by the underlying model, such as restricted diffusion and effects of diffusion time dependence (Jespersen et al., 2019, 2018).

In this study, we aim to more generally resolve the different kurtosis sources from the cumulant expansion of double diffusion encoding signals. Importantly, DDE signals can be used to extract information about the displacement correlation tensor ***Z*** (Jespersen, 2012; Jespersen et al., 2013; Jespersen and Buhl, 2011), which itself contains direct information about isotropic and anisotropic kurtosis contributions decoupled from intra-compartmental kurtosis effects. We additionally show that an index sensitive to intra-compartmental kurtosis arising from restricted diffusion can then be inferred by subtracting the total non-Gaussian information captured by the kurtosis tensor and the other inter-compartmental kurtosis source contributions. Given the prominence of the correlation tensor in this methodology, we term this approach Correlation Tensor Imaging (CTI). We present the theory underpinning CTI and provide its first experimental contrasts in *ex vivo* mouse brains and *in vivo* rat brains. While *ex vivo* experiments were designed to assess the full potential of CTI with high quality data acquired without scanning time constraints, the *in vivo* experiments were performed to demonstrate CTI's applicability and feasibility under *in vivo* conditions. Potential implications for future application of CTI are discussed.

.

# 2. Methods

2.1. Theory

2.1.1. Kurtosis Sources

Multiple Gaussian diffusion approximation: When the system consists of spins diffusing in non-exchanging microenvironments exclusively characterized by individual Gaussian diffusion tensors $\boldsymbol{D}^c$ (no restricted diffusion, no time dependence), the total signal measured from an SDE experiment can be described by the following equation (expressed with Einstein summation convention):

$$E_\Delta(\boldsymbol{q}) = \langle \exp[-q_i q_j \Delta D_{ij}^c] \rangle \qquad (1)$$

where $E_\Delta(\boldsymbol{q})$ is the normalized diffusion-weighted signal decay for a given q-vector $\boldsymbol{q}$ with magnitude $q = \gamma \delta g$ ($\gamma, \delta,$ and $g$ are the gyromagnetic ratio constant, gradient pulse duration and gradient intensity, respectively), $\Delta$ is the time interval between the gradient of a single diffusion encoding module (diffusion time), $D_{ij}^c$ are the elements of an individual diffusion tensor $\boldsymbol{D}^c$, and $\langle \cdot \rangle$ represents the average over different compartments (in biological tissues this will include any type of intra-cellular or extra-cellular contributions). In this study, the superscript 'c' is used to indicate properties of an individual component/compartment (n.b., in this study, we use "component" to reflect gaussian diffusion, "compartment" to imply restricted diffusion, and "microenvironment" is used quite loosely, i.e., it can reflect either components or compartments. None of the above is necessarily assigned to a particular biological component (e.g., intra/extra-cellular spaces)).

According to the standard DKI framework (Jensen et al., 2005), the total diffusion tensor $\boldsymbol{D}$ and total kurtosis tensor $\boldsymbol{W}$ of this system can be computed by decomposing equation 1 up to fifth order in $q$ (or up to the second order in $b = \Delta q^2$):

$$\log E_\Delta(\boldsymbol{q}) = -q_i q_j \Delta D_{ij} + \frac{1}{6} q_i q_j q_k q_l \Delta^2 \bar{D}^2 W_{ijkl} + \mathcal{O}(q^6) \qquad (2)$$

where $D_{ij}$ are the elements of the total diffusion tensor, $W_{ijkl}$ are the elements of the total kurtosis tensor, and $\bar{D}$ is the mean diffusivity (i.e., $\bar{D} = \text{trace}(\boldsymbol{D})/3$).

To factor out mesoscopic orientation dispersion of microenvironments (Callaghan et al., 1979; Jespersen et al., 2013; Kaden et al., 2016; Lasič et al., 2014), it is useful to compute the powder-averaged decays $\bar{E}_\Delta$ from the average of diffusion-weighted decays measured across q-vector samples $\boldsymbol{q}_i$ with different evenly distributed directions and constant magnitude $q$:

$$\bar{E}_\Delta(q) = \frac{1}{N_g} \sum_{i=1}^{N_g} E_\Delta(\boldsymbol{q}_i) \qquad (3)$$

where $N_g$ is the number of gradient directions. As for standard DKI (Jensen et al., 2005), equation 3 can be expanded in cumulants to extract the total excess-kurtosis $K_T$ of the powder-averaged signal decays (Henriques, 2018; Henriques et al., 2019; Westin et al., 2016):

$$\log \bar{E}_\Delta(q) = -\bar{D}\Delta q^2 + \frac{1}{6} K_T \bar{D}^2 \Delta^2 q^4 + \mathcal{O}(q^6) \qquad (4)$$

with

$$K_T = \frac{6}{5} \frac{\langle V_\lambda(\boldsymbol{D}^c) \rangle}{\bar{D}^2} + 3 \frac{V(\bar{D}^c)}{\bar{D}^2} \qquad (5)$$

where $\bar{D}^c$ is the mean diffusivity of individual tensors (i.e., $\bar{D}^c = \frac{\text{trace}(\boldsymbol{D}^c)}{3}$), $V_\lambda(\boldsymbol{D}^c)$ is the eigenvalue variance of an individual diffusion tensor $\boldsymbol{D}^c$, and $V(\bar{D}^c) = \langle \bar{D}^{c2} \rangle - \langle \bar{D}^c \rangle^2$ is the variance of mean diffusivities across microenvironments.

Equation 5 shows that, when microenvironments can be fully characterized by non-exchanging Gaussian diffusion components, $K_T$ can be fully decomposed by the anisotropic and isotropic kurtosis sources $K_{aniso}$ and $K_{iso}$:

$$K_T = K_{aniso} + K_{iso} \qquad (6)$$

with

$$K_{aniso} = \frac{6}{5}\frac{\langle V_\lambda(\boldsymbol{D}^c)\rangle}{\bar{D}^2} \qquad (7)$$

and

$$K_{iso} = 3\frac{V(\bar{D}^c)}{\bar{D}^2} \qquad (8)$$

It is important to note that when applying equation 2 to equations 4 and 5, the relationship between the total average excess-kurtosis $K_T$ and the kurtosis tensor $\boldsymbol{W}$ is given by the following expression:

$$K_T = \bar{W} + \Psi \qquad (9)$$

where $\bar{W}$ is the mean kurtosis tensor defined by (Hansen et al., 2016, 2013), i.e.:

$$\bar{W} = \frac{1}{5}(W_{1111} + W_{2222} + W_{3333} + 2W_{1122} + 2W_{1133} + 2W_{2233}) \qquad (10)$$

and $\Psi$ is a factor dependent on the mesoscopic orientation dispersion, that can be computed from the elements of the total diffusion tensor $\boldsymbol{D}$ yielding:

$$\Psi = \frac{2}{5}\frac{D_{11}^2 + D_{22}^2 + D_{33}^2 + 2D_{12}^2 + 2D_{13}^2 + 2D_{23}^2}{\bar{D}^2} - \frac{6}{5} \qquad (11)$$

<u>Intra-compartmental kurtosis effects</u>: Equation 6 does not consider non-Gaussian (restricted, time-dependent) diffusion arising from water molecules reflections on restricting boundaries (e.g., as can occur in intra- and extra-cellular compartments in tissues). This means that every component of the system is characterized only by its individual Gaussian diffusion (c.f. Eq.1).

To consider the restricted diffusion effects up to fourth order in $q$, it becomes necessary to add individual intra-compartmental kurtosis tensors $\boldsymbol{W}^c$ in equation 1 (Jespersen et al., 2019):

$$E_\Delta(\boldsymbol{q}) = \left\langle \exp\left[-q_i q_j \Delta D_{ij}^c + \frac{1}{6} q_i q_j q_k q_l \Delta^2 \overline{D}^{c2} W_{ijkl}^c \right] \right\rangle \quad (12)$$

where $W_{ijkl}^c$ are the elements of the individual intra-compartmental kurtosis tensor $\boldsymbol{W}^c$ arising from restricted or time-dependent diffusion. We also note that for compartments with a restricted diffusion-driven $\boldsymbol{W}^c$, $D_{ij}^c < D_0$ also reflects the time-dependent restricted diffusion effects on the diffusion tensor which are omitted for clarity in equation 12, $D_0$ is the free diffusion coefficient.

Applying this new representation to equations 3-4 and computing the cumulant expansion up to the fourth order in $q$, the total excess-kurtosis $K_T$ is now given by (Jespersen et al., 2019):

$$K_T = \frac{6}{5} \frac{\langle V_\lambda(\boldsymbol{D}^c) \rangle}{\overline{D}^2} + 3 \frac{V(\overline{D}^c)}{\overline{D}^2} + \frac{\langle (\overline{D}^c)^2 K_T^c \rangle}{\overline{D}^2} \quad (13)$$

where $K_T^c$ is the excess-kurtosis of powder averaged signals for an individual compartment "c", which can be computed as $(W_{1111}^c + W_{2222}^c + W_{3333}^c + 2W_{1122}^c + 2W_{1133}^c + 2W_{2233}^c)/5$. We note in passing that the individual $K_T^c$ arises solely from restriction, while $K_T$ emerges from all sources as explicitly shown in equation 13.

Equation 13 shows that, when non-Gaussian diffusion arises from reflection at boundaries, the apparent total excess-kurtosis $K_T$ encompasses contributions from the apparent isotropic and anisotropic kurtosis sources ($K_{aniso}$ and $K_{iso}$, c.f. equations 7 and 8) and contributions from individual excess-kurtosis $K_T^c$ averaged across all compartments – that will be referred to as the intra-compartmental kurtosis ($K_{intra}$) - i.e.:

$$K_T = K_{aniso} + K_{iso} + K_{intra} \quad (14)$$

with

$$K_{intra} = \frac{\langle (\overline{D}^c)^2 K_T^c \rangle}{\overline{D}^2}, \tag{15}$$

Note that $K_{intra}$ is not based on a specific microstructural model and thus it generally captures information from all restricting compartments, e.g., in white or gray matter tissues, whether arising, e.g., from intra-axonal, intra-cellular, or extra-cellular spaces. However, contributions from elements with higher mean diffusivity $\overline{D}^c$ will have contribute more to $K_{intra}$ as apparent from the numerator in equation 15.

2.1.2. Correlation tensor imaging

The correlation tensor imaging (CTI) framework is based on the correlation tensors in the cumulant expansion of double diffusion encoding (DDE) signals. Fig. 2A shows an illustration of the DDE gradient waveform and its parameters. DDE comprises two diffusion encoding modules characterized by different q-vectors ($\boldsymbol{q}_1$ and $\boldsymbol{q}_2$) and diffusion times ($\Delta_1$ and $\Delta_2$). The time interval between the two diffusion encoding modules is termed the mixing time ($\tau_m$). In this study, both diffusion times are equal, $\Delta_1 = \Delta_1 \equiv \Delta$.

Although, up to the fourth order in $q$, DDE signals can be related to a single 6th order kurtosis tensor (Hui and Jensen, 2015; Jensen et al., 2014), our correlation tensor imaging (CTI) approach is based on DDE's cumulant expansion formulated by Jespersen (2012) and the signal is expressed in terms of five unique second- and fourth-order tensors:

$$\log E_\Delta(\boldsymbol{q}_1, \boldsymbol{q}_2) = -(q_{1i}q_{1j} + q_{2i}q_{2j})\Delta D_{ij} + q_{1i}q_{2j}Q_{ij}$$

$$+ \frac{1}{6}(q_{1i}q_{1j}q_{1k}q_{1l} + q_{2i}q_{2j}q_{2k}q_{2l})\Delta^2 \overline{D}^2 W_{ijkl}$$

$$+ \frac{1}{4}q_{1i}q_{1j}q_{2k}q_{2l}Z_{ijkl}$$

$$+ \frac{1}{6}(q_{1i}q_{1j}q_{1k}q_{2l} + q_{2i}q_{2j}q_{2k}q_{1l})S_{ijkl} + \mathcal{O}(q^6)$$

$$\tag{16}$$

where, in addition to elements of the diffusion and kurtosis tensors ($D_{ij}$ and $W_{ijkl}$), $Q_{ij}$ are the elements of a 2nd order correlation tensor $\boldsymbol{Q}$ which encodes information on the time dependence of $\boldsymbol{D}$, and $Z_{ijkl}$ and $S_{ijkl}$ are the elements of the 4th order displacement correlation tensors $\boldsymbol{Z}$ and $\boldsymbol{S}$ (Jespersen, 2012).

2.1.3. Kurtosis separation using CTI

Anisotropic and isotropic kurtosis sources: At the long mixing time regime (and in the absence of flow and exchange (Jespersen, 2012)), $\boldsymbol{Z}$ becomes proportional to the diffusion tensor covariance $\boldsymbol{C}$ (Jespersen et al., 2013; Topgaard, 2017; Valiullin, 2017). Explicitly, $Z_{ijkl} \rightarrow 4\Delta^2 C_{ijkl}$ - which can then be used to generally estimate the ensemble averaged microscopic anisotropy $\langle V_\lambda(\boldsymbol{D}^c)\rangle$, and variance of mean diffusivities $V(\overline{D}^c)$ using the following equations (Jespersen et al., 2013; Topgaard, 2017; Valiullin, 2017):

$$\langle V_\lambda(\boldsymbol{D}^c)\rangle = \frac{2}{9}\big[C_{1111} + D_{11}^2 + C_{2222} + D_{22}^2 + C_{3333} + D_{33}^2 - C_{1122} - D_{11}D_{22} - C_{1133}$$
$$- D_{11}D_{33} - C_{2233} - D_{22}D_{33}$$
$$+ 3(C_{1212} + D_{12}^2 + C_{1313} + D_{13}^2 + C_{2323} + D_{23}^2)\big]$$

(17)

and

$$V(\overline{D}^c) = \frac{1}{9}(C_{1111} + C_{2222} + C_{3333} + 2C_{1122} + 2C_{1133} + 2C_{2233}) \qquad (18)$$

It is important to note that equations 17 and 18 do not rely on the multiple Gaussian diffusion assumption and thus, in the absence of flow and exchange, $\langle V_\lambda(\boldsymbol{D}^c)\rangle$ and $V(\overline{D}^c)$ can be generally converted to the anisotropic and isotropic kurtosis sources using equations 7 and 8.

Intra-compartmental kurtosis index: Given that the total excess-kurtosis $K_T$ can be obtained from $\boldsymbol{W}$ and $\boldsymbol{D}$ (eq. 9) and that the two inter-compartmental kurtosis sources $K_{aniso}$ and $K_{iso}$ can be derived from **Z** and **D** (eqs. 7-8, 17-18), an index sensitive to the intra-compartmental kurtosis can be computed from the following simple subtraction:

$$K_{intra} = K_T - K_{aniso} - K_{iso} \qquad (19)$$

since $K_T$ contains the information on all three kurtosis sources.

<u>Higher order terms:</u> Note that, since CTI is based on the truncated cumulant expansion of DDE signals, the accuracy of all indices, and especially the $K_{intra}$ index, may be compromised by the higher order effects neglected in the above expressions (c.f. $\mathcal{O}(q^6)$ in Eq. 16), particularly if the data are not acquired at sufficiently low b-values (Chuhutin et al., 2017; Ianuş et al., 2018). In this study, we only explored these higher order effects empirically (i.e., by probing b-value dependencies and using simulations to provide intuition). Exploring ways to minimize these effects was beyond the scope of this study and will be reported in due course.

### 2.1.4. Acquisition requirements for CTI

<u>Suppressing $Q_{ij}$ and $S_{ijkl}$:</u> As shown above, only **D**, **W** and **Z** are necessary to resolve the contributions of the different kurtosis sources. DDE has an appealing "built in" suppressor of the **Q** and **S** tensors via the mixing time: at long $\tau_m$, both tensors vanish (Jespersen, 2012). Alternatively, these tensors can be cancelled out by combining DDE acquisitions with inverted $\boldsymbol{q}_2$ vectors:

$$\frac{\log E_\Delta(\boldsymbol{q}_1, \boldsymbol{q}_2)}{2} + \frac{\log E_\Delta(\boldsymbol{q}_1, -\boldsymbol{q}_2)}{2} =$$

$$-(q_{1i}q_{1j} + q_{2i}q_{2j})\Delta D_{ij}$$

$$+\frac{1}{6}(q_{1i}q_{1j}q_{1k}q_{1l} + q_{2i}q_{2j}q_{2k}q_{2l})\Delta^2 \overline{D}^2 W_{ijkl}$$

$$+\frac{1}{4}q_{1i}q_{1j}q_{2k}q_{2l}Z_{ijkl} + \mathcal{O}(q^6)$$

(20)

To reduce the number of parameters to be fitted, the contributions of tensors $\boldsymbol{Q}$ and $\boldsymbol{S}$ were here suppressed in this way.

<u>Diffusion-weighting requirements:</u> Analogously to Diffusional Kurtosis Imaging (DKI), fitting tensors associated with the $q^4$ cumulant (i.e. $\boldsymbol{W}$ and $\boldsymbol{Z}$), CTI requires data acquired with at least three different diffusion gradient intensities (shells). In addition, asymmetric DDE gradients are required for decoupling of elements $W_{ijkl}$ and $Z_{ijkl}$. In this study, a minimal protocol for CTI was designed based on the following eight gradient intensity combinations of $q \in \{0, q_a, q_b\}$ (Fig. 2B):

1) $|\boldsymbol{q}_1|, |\boldsymbol{q}_2| = q_a, q_a$;

2) $|\boldsymbol{q}_1|, |\boldsymbol{q}_2| = q_a, q_a$, with inverted $\boldsymbol{q}_2$ direction;

3) $|\boldsymbol{q}_1|, |\boldsymbol{q}_2| = q_a, 0$;

4) $|\boldsymbol{q}_1|, |\boldsymbol{q}_2| = 0, q_a$;

5) $|\boldsymbol{q}_1|, |\boldsymbol{q}_2| = q_b, q_b$;

6) $|\boldsymbol{q}_1|, |\boldsymbol{q}_2| = q_b, q_b$, with inverted $\boldsymbol{q}_2$ direction;

7) $|\boldsymbol{q}_1|, |\boldsymbol{q}_2| = q_b, 0$; and

8) $|\boldsymbol{q}_1|, |\boldsymbol{q}_2| = 0, q_b$.

The magnitudes $q_a$ and $q_b$ were defined for a given $b_{max}$ value ($q_a = \sqrt{b_{max}/2(\Delta - \delta/3)}$ and $q_b = \sqrt{b_{max}/(\Delta - \delta/3)}$).

<u>Gradient direction requirements:</u> To resolve the anisotropic information of $\boldsymbol{Z}$, different pairs of gradient directions are required (Jespersen et al., 2013). The orientations defined in the DDE 5-design (Jespersen et al., 2013) are adequate for this purpose, namely, twelve pairs of parallel $\boldsymbol{q}_1$-$\boldsymbol{q}_2$ directions and sixty pairs of perpendicular $\boldsymbol{q}_1$-$\boldsymbol{q}_2$ directions. In this study, to decrease the difference between the number of parallel and perpendicular $\boldsymbol{q}_1$-$\boldsymbol{q}_2$ directions, 45 extra DDE experiments with parallel $q_1$-$q_2$ directions were acquired, yielding a total of 117 (57 parallel + 60 perpendicular) $q_1$-$q_2$ combination of directions. The parallel directions of these latter $q_1$-$q_2$ parallel pairs were evenly sampled on a spherical 3-dimensional grid. All 117 direction combinations were repeated for the $q_1$-$q_2$ magnitude combinations 1-8 described above.

2.2. MRI experiments

All animal experiments were preapproved by the competent institutional and national authorities, and carried out according to European Directive 2010/63.

*Ex vivo* CTI experiments. Brain specimens were extracted via transcardial perfusion with 4% Paraformaldehyde (PFA) from two adult mice (N=2 C57BL/6J males 13 weeks old, weights 23/24g, respectively, grown with a 12h/12h light/dark cycle with *ad libitum* access to food and water). After extraction from the skull, both brains were immersed in 4% PFA solution for 24 h, and then washed in Phosphate-Buffered Saline (PBS) solution for at least 48 h. The specimens were then placed in a 10-mm NMR tube filled with Fluorinert (Sigma Aldrich, Lisbon, PT), secured with a stopper from above to prevent floating, and the NMR tube was sealed using paraffin film. MRI scans were prefomed on a 16.4 T Aeon Ascend Bruker scanner (Karlsruhe, Germany) equipped with an AVANCE IIIHD console, and a Micro5 probe with gradient coils capable of producing up to 3000 mT/m in all directions. Using the probe's variable temperature capability, we maintained the samples at 37$_o$C. The samples were allowed to equilibrate with the surroundings for at least 3 h prior to commencement of diffusion MRI experiments.

Double diffusion encoding data was acquired for five coronal slices using an in house written EPI-based DDE pulse sequence. The diffusion encoding gradient pulse separation Δ and mixing time $\tau_m$ were set to 13 ms, while the pulsed gradient duration δ was set to 1.5 ms (Fig. 2A). Data were acquired for the minimal CTI protocol using the 117 DDE pairs of directions and repeated for all eight $\boldsymbol{q}_1$-$\boldsymbol{q}_2$ magnitude combinations (parameters described in section 2.1.4), in addition to sixty acquisitions without any diffusion-weighted sensitization (zero b-value). To assess the effects of higher order terms not considered by CTI, the entire set of minimal protocol acquisitions were repeated for seven evenly sampled $b_{max}$ values (1.00, 1.25, 1.50, 1.75, 2.00, 2.25, and 2.5 ms/μm2). For all experiments, the following common parameters were used: TR/TE = 2200/52 ms, Field of View = 10.4 × 10.4 mm2, matrix size 80 × 80, leading to an in-plane voxel resolution of 130×130 μm2, slice thickness = 0.9 mm, slice gap = 0.6 mm, number of segments = 2, number of averages = 8, partial fourier effective acceleration = 1.42. For each $b_{max}$ value, the total acquisition time was about 9.75 h.

In addition to the diffusion-weighted data, coronal T2-weighted images with high resolution and high SNR were acquired for anatomical reference and assessment of partial volume effects. This data was acquired using a RARE sequence with the following parameters: TR = 4250 ms, effective TE = 22 ms, RARE factor = 8, Field of View = 10 × 10 mm2, matrix size 126 × 126, in-plane voxel resolution = 79.4 × 79.4 μm2, slice thickness = 79.4 μm (40 slices), number of averages = 230, partial fourier effective acceleration = 1.05.

*In vivo* CTI experiments. To assess CTI's ability to characterize tissues *in vivo* within feasible scan times, data was acquired on two rats *in vivo* (N=2 Long Evans females, 14/15 weeks old, weights 264/254 g, respectively, grown in a 12h/12h light/dark cycle with *ad libitum* access to food and water) under anesthesia (Isoflurane ~2.5% in 28% oxygen). All *in vivo* data were acquired on a 9.4 T Bruker Biospec MRI scanner equipped with an 86 mm quadrature coil for transmission and a 4-element array cryocoil for reception. Due to acquisition time constraints, *in vivo* double diffusion encoding data were acquired only for one CTI minimal protocol with $b_{max}$ = 2 ms/µm2 ($\Delta$ = 12 ms, $\tau$ = 12 ms, $\delta$ = 3 ms, 117 pairs of direction for 8 gradient intensity combinations, c.f. Fig. 2B) for three coronal images using per slice respiratory gating. A large number of 180 b-values = 0 data were also acquired to ensure a high ratio between the number of non-diffusion and diffusion-weighted acquisitions (Alexander and Barker, 2005; Jones et al., 1999)). Other acquisition parameters included: TR/TE = 3000 / 48.5 ms, Field of View = 20 × 20 mm2, matrix size 100 × 100, in-plane resolution of 200 × 200 µm2, slice thickness = 1 mm, slice gap = 1.8 mm, number of EPI segments = 1, number of averages = 2, partial fourier effective acceleration = 1.40. For each animal, the acquisition time of all of the diffusion data was about 2 hours.

Data processing: All diffusion-weighted datasets were first preprocessed by realigning the data using a sub-pixel registration technique (Guizar-Sicairos et al., 2008). CTI was then directly fitted to the data using a weighted-linear-least squares fitting procedure, implemented in-house, to equation 20. No terms associated with cumulants higher than the fifth order in $q$ were considered in our fitting procedure to ensure the precision of kurtosis estimates (Kiselev, 2017). Two different analyses were performed for each *ex vivo* mouse brain dataset: (i) to assess the contrasts of different kurtosis sources with maximum precision, CTI estimates were obtained in a single fit incorporating all diffusion-weighted datasets of all $b_{max}$ values, making a total of 56 combination of $\boldsymbol{q}_1$-$\boldsymbol{q}_2$ magnitudes and 420 b-value = 0 acquisitions; (ii) CTI estimates were additionally produced for individual $b_{max}$ sets in the minimal protocols acquired, to test the robustness of kurtosis estimates with different b-values and to assess the effects of higher order terms. For the *in vivo* acquisitions, CTI estimates of each rat brain dataset was processed for the single acquired minimal protocol defined with a $b_{max}$ = 2 ms/µm$^2$. In addition to visual inspection of the CTI derived maps, kurtosis estimates were also extracted from regions of interest (ROIs) which were bilateratly drawn on all data slices.

## 2.3. Simulations

To support the interpretation of the results from the MRI experiments, the CTI approach was also subjected to extensive numerical simulations using noise-free synthetic signals in which ground truth kurtosis sources are known *a priori* (simulations with added synthetic noise are reported in supplementary material). The simulations were performed using the minimal CTI protocol defined by the eight gradient intensity combinations decribed in Fig. 2B, which were repeated for the gradient directions of Jespersen's 5-design in addition to 45 extra parallel directions. It is important to note that the eight gradient intensity combinations decribed in Fig. 2B are defined for a given $b_{max}$ value, and thus apparent kurtosis estimates can be obtained as a function of $b_{max}$. To assess the effects of higher order terms, kurtosis estimates were independently produced for seven different $b_{max}$ values (i.e., 1.00, 1.25, 1.50, 1.75, 2.00, 2.25, and 2.5 ms/μm2). The diffusion gradient separation times, pulse durations and mixing time for these simulations were set to Δ = 13 ms, δ = 1.5 ms, and $\tau_m$ = 13 ms, respectively (the same parameters used in the *ex vivo* mouse brain experiments). The synthetic signals were produced according to two different ground-truth scenarios comprising a mix of Gaussian components and a mix of Gaussian and/or restricted compartments:

1) <u>Gaussian microenvironements according to a two-component model:</u> DDE signals were first produced for two well-aligned axially-symmetric Gaussian diffusion components (no restricted diffusion). The axial and radial diffusivities for the first component were set to 2 and 0 μm2/ms, while the axial and radial diffusivities for the second component were set to 1.5 and 0.5 μm2/ms. Volume fractions for both components were set to 0.5. Based on these values, ground truth kurtosis was computed using equations 7 and 8. Note that for these simulations $K_{intra} = 0$. To assess the dependence of the signals on mesoscopic orientations, simulations were also repeated for different levels of orientation dispersion.

For this, simulations of 10000 replicas of the symmetric Gaussian diffusion components were produced. The directions of these replicas were sampled based on a Watson distribution (Watson, 1965) which can be produced with arbitrary dispersion levels. For this study, different dispersion levels were tested by changing the Watson distribution concentration parameter $\kappa$ from 0 to 16.58, where $\kappa = 0$ corresponds to completely randomly oriented microenvironments, while $\kappa = 16.58$ corresponds to a low dispersion of 10 degrees (according to the dispersion angle definition proposed by (Riffert et al., 2014)).

2) <u>Gaussian component and restricted microenvironements:</u> synthetic DDE signals for non-zero $K_{intra}$ were produced by incorporating a spherical compartment (restricted diffusion) to the two Gaussian components described above. The signals for the restricted compartment were produced using the MISST package (Drobnjak et al., 2011). Simulations were repeated for three different sphere diameters ($d_i$ = 5, 7.5, and 10 μm, approximately mimicking typical cell soma sizes in the nervous system) and the volume fractions for both Gaussian components and for the restricted compartment were set to have equal contributions (i.e. $f_1 = f_2 = f_3 = 1/3$). Other simulation parameters were as follows: intrisic diffusivity = 2 μm2/ms; simulations sampling time = 0.015 ms; number of spherical Bessel orders = 70 (Callaghan, 1997; Drobnjak et al., 2011). The ground truth individual mean diffusivity $\overline{D}^c$ and individual excess-kurtosis $\overline{K}^c$ of the spherical restricted compartment were determined using extra SDE simulations. To achieve the apparent values for $q \to 0$, these ground truth values were computed by fitting the standard diffusional kurtosis imaging (DKI) equation (Jensen et al., 2005) to the synthetic signals simulated for a maximum gradient intensity set to $q_{max} = 0.1/d_i$. To avoid fitting instabilities, SDE synthetic signals were evenly sampled for 2500 b-values from 0 to $b_{max} = (2\pi q_{max})^2(\Delta - \delta/3)$. The ground truth $K_{aniso}$, $K_{iso}$ and $K_{intra}$ for the total synthetic signals were computed using equations 7, 8, and 15. These simulations were also repeated for different levels of orientation dispersion of the axially-symmetric compnents. For this, 10000 replicas of the Gaussian and spherical compartments were sampled based on a Watson distribution with a varying dispersion level. For the sake of simplicity, simulations of restricted spherical compartments and dispersed Gaussian compartments were only produced for the larger sphere diameter of 10 μm, which corresponds to the scenario with larger $K_{intra}$.

# 3. Results

3.1. MRI experiments

Raw data of *ex vivo* DDE-MRI experiments for different b-values are presented in Fig. 3. ROIs placed in white matter (WM), grey matter (GM) and cerebral ventricles (CV) at b = 0 (Fig. 3A) exhibited signal-to-noise ratios (SNRs) of 70±20, 110±10, and 192±2 for all WM, GM, and CV ROIs, respectively. For a representative slice, Fig. 3B-D shows the powder-averaged DDE signal decays at total b-value of 1, 3 and 5 ms/µm2, respectively, and for parallel DDE experiments ($\bar{E}_\Delta(\boldsymbol{q}_1 \uparrow\uparrow \boldsymbol{q}_2)$, Fig. 3 B1, C1, D1), anti-parallel DDE experiments ($\bar{E}_\Delta(\boldsymbol{q}_1 \uparrow\downarrow \boldsymbol{q}_2)$, Fig. 3 B2, C2, D2), and perpendicular DDE experiments ($\bar{E}_\Delta(\boldsymbol{q}_1 \perp \boldsymbol{q}_2)$, Fig. 3 B3, C3, D3), respectively. At these b-values, individual diffusion-weighted images were characterized by SNRs of 41±9, 19±1, and 10±1 for all WM ROIs, and SNRs of 62±4, 20±1, and 8±1 for all GM ROIs (the SNR values are from powder-averaged parallel DDE experiments, Fig. 3 B1, C1, D1).

An important assumption of CTI, as presented here, is that the long mixing time regime has been achieved. This can be empirically tested by comparing data acquired with parallel and anti-parallel diffusion pairs. Maps corresponding to the signal ratios between parallel and anti-parallel DDE measurements are shows in panels B4, C4, and D4 of Fig. 3. The ratio maps show values near unity, indicating that the long mixing time regime assumption ($\lim_{\tau \to \infty} \bar{E}_\Delta(\boldsymbol{q}_1 \uparrow\uparrow \boldsymbol{q}_2) / \bar{E}_\Delta(\boldsymbol{q}_1 \uparrow\downarrow \boldsymbol{q}_2) \to 1$, (Jespersen and Buhl, 2011; Koch and Finsterbusch, 2008; Ozarslan, 2009)) is practically fulfilled for most voxels. For comparison, the maps of the ratio between parallel and perpendicular DDE signals ($\bar{E}_\Delta(\boldsymbol{q}_1 \uparrow\uparrow \boldsymbol{q}_2) / \bar{E}_\Delta(\boldsymbol{q}_1 \perp \boldsymbol{q}_2)$) are shown in panels B5, C5, and D5 of Fig. 3. These reveal the expected higher values for white matter regions where microscopic anisotropy is known to be higher. As predicted by (Ianuş et al., 2018; Jespersen et al., 2013), this contrast increased with higher b-value (Fig. 3B5, Fig. 3C5, and Fig. 3D5).

Given the robustness of the raw data and the fulfillment of the long mixing time regime, the correlation tensor metrics were first extracted from the extensively sampled b-value protocol (i.e. data were fit using all 56 acquired $\boldsymbol{q}_1$-$\boldsymbol{q}_2$ magnitude combinations together). Fig. 4 presents the kurtosis source separation maps for each of the five slices acquired for a representative mouse brain. $K_T$, $K_{aniso}$, $K_{iso}$, and $K_{intra}$ estimates (shown in panels A-D, respectively) evidenced drastically different contrasts. Notably, $K_T$ is, as expected, higher than any of its sources (Fig. 4A). $K_{aniso}$ is revealed to be the largest source contributing to the total kurtosis in white matter (e.g. regions pointed by white arrows, Fig. 4B). On the other hand, $K_{iso}$ shows relatively low intensities for both white and grey matter, with the exception of areas where partial volume effects arising from free water in cerebral ventricles are dominant (e.g. regions pointed by grey arrows, Fig. 4C). Large partial volume effects in these areas are supported by high-resolution anatomical mapping (supplementary Fig. S1), that directly showed areas with free water overlapping with other tissues. $K_{intra}$ maps mainly exhibited positive values (25th and 75th percentiles for all data voxels were 0.286 and 0.430, respectively). Relative to $K_{aniso}$, both $K_{iso}$ and $K_{intra}$ maps seem to be more sensitive to image artefacts (e.g. dorsal-ventral oscillations pointed by the red arrows, which are likely to be a combined effect of Gibbs Ringing and partial Fourier). All these results were found to be consistent between the N=2 *ex vivo* mouse brains scanned, as shown in supplementary Fig. S2. For instance, $K_{intra}$ maps for mouse brain #2 show consistent positive values (25th and 75th percentiles for all data voxels are 0.320 and 0.431, respectively).

Kurtosis estimates extracted for individual $b_{max}$ minimal protocols are shown in Fig. 5, namely, $K_T$, $K_{aniso}$, $K_{iso}$, and $K_{intra}$ maps are displayed for both mice. The kurtosis mean values and standard deviation across the animals are plotted as a function of the protocol values $b_{max}$ in Fig. 5A3-D3. For reference, the mean kurtosis estimates of the extensively sampled b-value protocol are plotted (dotted lines) in these latter panels. Across the two mouse brain specimens, $K_T$ maps are consistent only for the higher b-values and evidence implausible negative values at low b-values (Fig. 5A1-2). $K_{aniso}$ estimates decreased as $b_{max}$ increased (Fig. 5B) – this b-value dependence can be particularly appreciated by observing the $K_{aniso}$ estimates extracted from the white matter ROIs (Fig. 5B3). Similarly to the $K_T$ maps, $K_{iso}$ maps were visually nosier at lower b-values (Fig. 5C1 and Fig. 5C2). The b-value dependence for mean $K_{iso}$ is visually less obvious than the other kurtosis estimates (Fig. 5C3). $K_{intra}$ maps consistently exhibited positive values at the higher b-values; however, negative values were present in the noisier $K_{intra}$ maps at lower $b_{max}$ values (Fig. 5D).

Results from the *in vivo* rat experiments are shown in Fig. 6. White matter and grey matter ROIs placed in the superior regions of the rat brain (green and red ROIs in Fig. 6) exhibited, respectively, SNRs of 41 ± 5 and 41 ± 4 for the non-diffusion weighted data, SNRs of 12 ± 2 and 10 ± 2 for total b-value = 2 ms/µm2, and SNRs of 5.3 ± 0.9 and 3.1 ± 0.4 for total b-value = 4 ms/µm2. Inferior areas of the brain in our *in vivo* data (regions associated with a lower cryocoil sensitivity) showed lower signal intensities and thus lower SNRs, particularly, a grey matter ROIs placed in the inferior region of the brain (blue ROI in Fig. 6A) evidenced SNRs of 26 ± 4, 6.2± 0.8, and 2.4± 0.4 for the total b-values of 0, 2 and 4 ms/µm2, respectively. $K_T$, $K_{aniso}$, $K_{iso}$, and $K_{intra}$ maps for all slices and for both animals are shown in Fig. 6 B-E. Note that different kurtosis types are displayed with different color bar ranges for better contrast visualization. Consistent with the *ex vivo* CTI results, $K_T$ was higher than any of its sources (Fig. 6B) and $K_{aniso}$ dominated in white matter (Fig. 6C). *In vivo* $K_{iso}$ maps also demonstrated the abovementioned sensitivity towards partial volume effects between cerebral tissue and cerebral ventricle (free water pointed by black arrows), but also appeared higher in WM (white arrows) (Fig. 6D). High $K_{iso}$ estimates were also present in the grey matter of inferior brain regions (pointed by red arrows). For both *in vivo* rats, $K_{intra}$ was consistently positive (Fig. 6E, 25th-75th percentile range of all slice voxels is 0.21-0.45 for Rat 1 and 0.18-0.43 for Rat 2).

3.2. Numerical simulations

To further validate CTI and investigate how the different sources of kurtosis would vary with b-value, numerical simulations for several plausible diffusion conditions were performed. When the system consists of Gaussian components, namely, perfectly aligned "sticks" (with zero radial diffusivity) and a tensor (Fig. 7A), $K_T$ and $K_{aniso}$ extracted from CTI shows a weak dependence on b-value; in particular, the biases introduced by higher order effects are lower than 1.2% (Fig. 7A1-2). On the other hand, the apparent $K_{iso}$ extracted from CTI appears lower with increasing b-value due to higher order terms, reaching negative biases larger than 50% at $b_{max} = 2.5$ ms/µm2 (Fig. 7A3). In this system, $K_{intra}$ ground truth is identically zero; however, higher order terms induce a positive apparent $K_{intra}$ (maximum bias of ~0.01 for $b_{max} = 2.5$ ms/µm2, Fig. 7A4). When orientation dispersion is added to the same system (Fig. 7B), the extracted parameters have different dependencies on b-value. $K_T$ and $K_{aniso}$ extracted from CTI approach their nominal ground truth only at b = 0, while they are increasingly underestimated at higher b-values (Fig. 7B1-2). $K_{iso}$ and $K_{intra}$ extracted from CTI now become increasingly overestimated with higher b-values (Fig. 7B3-4). Particularly, $K_{intra}$ bias becomes positive, higher than 0.2 for the larger $b_{max}$, when Gaussian components are completely randomly oriented (yellow curve of Fig. 7B4)

To assess how intra-compartmental restricted diffusion may affect the b-value dependence of the extracted parameters, simulations incorporating an impermeable sphere along with the stick and tensor model above, were performed (Fig. 8). Without orientation dispersion (Fig. 8A), $K_T$, $K_{aniso}$ and $K_{iso}$ values approach their ground-truth at low b-values but are all underestimated at higher b-values (Fig. 8A1-3). Although large positive bias is observed at higher b-values, apparent $K_{intra}$ values are negative for low b-values as expected for the negative concave signal decays profiles of restricted diffusion (Fig. 8A4). $K_{intra}$ bias trends depend on the size of the sphere, yet apparent $K_{intra}$ values approached their respective negative ground truth values at low b-values (Fig. 8A4). Fig. 8B shows the simulations when a restricted sphere of 10 μm is added to orientationally-dispersed replicas of the two-Gaussian compartments; the trends are similar to those reported above, although the underestimation of $K_{intra}$ becomes higher in this scenario (Fig. 8B4).

# 4. Discussion

Since its inception, DKI has played an important role in microstructural characterizations. In many cases, kurtosis measurements appeared more sensitive to disease or other processes, such as development and ageing, compared with their diffusion tensor counterparts (Cheung et al., 2012; Falangola et al., 2008; Henriques, 2018; Lin et al., 2018; Rudrapatna et al., 2014; Sun et al., 2015). Clinical applications of diffusional kurtosis MRI abound, and deeper investigations into kurtosis features, such as its time dependence, are being vigorously studied (Grussu et al., 2019; Jensen and Helpern, 2010; Jespersen et al., 2018; Lee et al., 2019; Pyatigorskaya et al., 2014). Nearly invariably, these measurements are performed using single diffusion encoding pulse sequences; however, SDE methods cannot separate the different sources of kurtosis, which reduces the specificity of DKI. QTE approaches have recently been gaining much interest for their ability to portray anisotropic and isotropic kurtosis sources separately (Lasič et al., 2014; Szczepankiewicz et al., 2019, 2016, 2015; Topgaard, 2019, 2017; Westin et al., 2016); however, the strong model assumptions underlying QTE may limit the confidence in the specificity and the interpretation of the extracted kurtosis sources (Jespersen et al., 2019). Specifically, time dependent and intra-compartmental diffusional kurtosis effects may corrupt QTE estimates in a non-linear way. Other recent studies have attempted to decouple sources of inter- and intra-compartmental kurtosis using a frequency modulation of specific symmetrized DDE experiments (Ji et al., 2019; Paulsen et al., 2015). Nevertheless, that approach is confounded by orientation dispersion (Paulsen et al., 2015), i.e., the frequency modulation in these symmetrized DDE experiments conflates the different sources of kurtosis: intra-compartmental kurtosis, anisotropic kurtosis, and orientation dispersion.

## 4.1. CTI theoretical potentials and experimental requirements

In this study, we sought to develop a methodology capable of resolving anisotropic kurtosis ($K_{aniso}$), isotropic kurtosis ($K_{iso}$), and intra-compartmental kurtosis ($K_{intra}$) for any number of non-exchanging tissue compartments (in the absence of flow) and without relying on the assumption of Gaussian (or multiple Gaussian) diffusion. Our correlation tensor imaging (CTI) approach allows the estimation of the correlation tensor $\boldsymbol{Z}$, which is expressed in DDE but not SDE signals. We have demonstrated, as previously predicted, that the $\boldsymbol{Z}$ tensor can provide the sought-after information, provided that the long mixing time regime is reached (Jespersen, 2012; Jespersen et al., 2013) so that the contributions of the $\boldsymbol{Q}$ and $\boldsymbol{S}$ tensors are approximately zero (mixing time approximation #1) and the $\boldsymbol{Z}$ tensor is approximately equal to $4\Delta^2\boldsymbol{C}$ (mixing time approximation #2). Although the signals arising from the $\boldsymbol{Q}$ and $\boldsymbol{S}$ tensors can be eliminated by acquiring DDE experiments with an inverted gradient direction (c.f. Eq. 20), reaching the long mixing time is still necessary to fulfil the mixing time approximation #2 such that the anisotropic and isotropic kurtosis sources can be accurately extracted from the $\boldsymbol{Z}$ tensor (c.f. Eqs. 17 and 18). The long mixing time regime need not be an implicit assumption of CTI since this regime can be empirically identified by comparing DDE signals with parallel and antiparallel experiments (Jespersen and Buhl, 2011; Lawrenz and Finsterbusch, 2013; Ozarslan, 2009), as indeed done in this study. Since both mixing time approximations are associated with the same characteristic length, observing that the signals from parallel and antiparallel experiments are identical (i.e., their ratios are near unity) is likely to be a good indication that both $\boldsymbol{Q}, \boldsymbol{S} \rightarrow 0$ and that $\boldsymbol{Z} \approx 4\Delta^2 \boldsymbol{C}$. Our finding that the ratio between parallel and antiparallel experiments is close to unity in the entire brain (Fig. 3), already at a mixing time of 13 ms, is consistent with previous studies suggesting the long mixing time is reached in neural tissues rather rapidly (Henriques et al., 2019; Ianuş et al., 2018; Shemesh et al., 2012, 2011; Shemesh and Cohen, 2011) and consistent with our simulations which showed that $K_{aniso}$ and $K_{iso}$ estimates are close to their ground truth values (at least for $b_m \rightarrow 0$).

To map the correlation tensor directly, typical DDE experiments with $|\boldsymbol{q}_1| = |\boldsymbol{q}_2|$ are insufficient. Although such experiments can resolve $K_{aniso}$ and other microscopic anisotropy measures, e.g. (Ianuş et al., 2018; Jespersen et al., 2013), asymmetric DDE wavevector magnitudes are necessary for decoupling the full set of elements in the $\boldsymbol{Z}$ tensor from the $\boldsymbol{W}$ tensor (c.f. Eq. (24)) and for measuring $K_{iso}$ and $K_{intra}$. An efficient way to construct an asymmetric DDE protocol is to incorporate measurements whereby one of the DDE wavevectors is set to zero, effectively making the acquisition scheme a combination of SDE and DDE measurements (linear and planar encoding). However, we note that setting one of the wavevectors to zero is not necassary and in other applications, finite yet unequal magnitudes may be beneficial. In future studies, the CTI acquisition protocol could be further optimized by exploring other b-value and gradient direction combinations potentially better suited for extraction of different kurtosis sources or for accelerating acquisition times.

We note that although the mathematical framework of CTI was derived based on the assumption that tissue can be represented by a number of non-exchanging signal contributions, exchange across components can only decrease the kurtosis values measured here (Fieremans et al., 2010; Jensen et al., 2005; Jensen and Helpern, 2010; Ning et al., 2018).

### 4.2. Sources of non-Gaussian diffusion in brain tissues

Fitting the *ex vivo* dataset with heavily sampled b-values yielded robust maps of kurtosis sources in the mouse brain (Fig. 4). The usefulness of CTI can be gleaned by noticing that $K_T$ is similarly high in white matter (white arrows) and in areas near the ventricles (grey arrow). Our CTI approach resolved different sources of kurtosis in these two areas: in white matter, $K_{aniso}$ dominates the high excess-kurtosis, while in the areas close to the ventricles, $K_{iso}$ underlies the high excess-kurtosis, due to a wide distribution of diffusivities arising from the large partial volume effects between tissue water and freely diffusing water in the cerebral ventricles. In deep white matter and grey matter regions, $K_{iso}$ is the smaller kurtosis source, which is consistent with recent studies showing that isotropic diffusion encoding signals exhibit smaller deviations from mono-exponential decay in similar areas (Dhital et al., 2018; Szczepankiewicz et al., 2015).

Theoretically, negative $K_{intra}$ measurements would be expected for the concave signal decay arising from fully restricted diffusion, that in some circumstances can lead to diffusion diffraction effects at high b-values (Avram et al., 2004; Callaghan, 1995; Callaghan et al., 1991; Cohen and Assaf, 2002). On the other hand, positive $K_{intra}$ (scalar) values may be expected for convex signal decays arising from diffusion in compartments with variable cross-section dimensions, or in the presence of sub-structures (Dhital et al., 2018; Novikov and Kiselev, 2010). Vanishing $K_{intra}$ could be expected at very long diffusion time regimes (Jespersen et al., 2007; Novikov et al., 2018a). For systems comprising an ensemble of different compartment types, the observed $K_{intra}$ index corresponds to a weighted sum of all intra-compartmental kurtosis sources (c.f. Eq. 15), where compartments with higher diffusivities have higher weights (e.g., perhaps diffusion in extra-cellular spaces). In this study, CTI's $K_{intra}$ measurements from both *ex vivo* and *in vivo* experiments consistently evidenced positive values ranging from ~0.3 to ~0.6. Although these positive values seem to suggest that positive intra-compartmental kurtosis sources prevail in brain tissues, it is important to stress that our simulations suggest that $K_{intra}$ estimates measured with the current protocol are likely to be compromised by high order terms not accounted for by CTI (*vide infra*). Although our results in general revealed consistent positive values across animals (both *ex vivo* and *in vivo*), it is currently difficult to infer more subtle details on how $K_{intra}$ varies along different brain areas due to the small dimensions of the mice/rat brain structures and due to small differences in slice positioning and orientation across the animals, in addition to the bias from higher order terms. Further studies are thus required for asserting the nature of intra-compartmental kurtosis.

### 4.3. Trade-off between accuracy and precision

Like all kurtosis measurements, higher order terms can play an important role in the accuracy and precision of the estimated metrics (Chuhutin et al., 2017; Ianuş et al., 2018). Since kurtosis is formally defined at b-value = 0, it (and any other metric defined from the cumulant expansion) is inherently biased when measured at any finite b-value. Similarly, a biased metric will be measured in the CTI framework. We thus sought to explore the impacts of higher order effects by examining CTI's b-value dependence in a controlled way – namely, by varying $b_{max}$ in the minimal protocol (c.f. Fig. 5). The higher order terms are expected to have a larger impact on parameter accuracy at higher b-values; however, lower b-values might not provide sufficient diffusion-weighting for precise kurtosis estimation. Indeed, our results show that kurtosis estimates are not qualitatively consistent across e*x vivo* mouse brain specimens for $b_{max}$ values lower than 1 ms/μm2, and thus, we suggest that a $b_{max}$ value between 1.5 and 2 ms/μm2 may provide, in practice, an optimal trade-off between precision and accuracy. Kurtosis estimates for higher $b_{max}$ vales (2.25 and 2.5 ms/μm2, c.f. Fig. 5) indicate that higher order terms might introduce negative biases in $K_{aniso}$ and positive biases in $K_{intra}$. $K_{iso}$ estimates exhibited both positive and negative bias trends depending on the specific brain regions – a feature which should be considered in future studies.

Our simulations confirmed the expected trends between CTI's accuracy and precision (c.f., Fig. 7, Fig 8, Fig. S3, and Fig. S4): firstly, negative bias trends for synthetic $K_{aniso}$ estimates and positive bias trends for synthetic $K_{intra}$ estimates were observed for orientation dispersion and/or non-Gaussian effects from restricted compartments (Fig. 7B, Fig. 8A, and Fig. 8B); secondly, our simulations confirmed that the signs of $K_{iso}$ bias can vary with different microstructural scenarios. Particularly, pronounced higher order effects in $K_{iso}$ are expected in regions containing partial volume effects with free water, since free water components have a strong signal attenuation due to high diffusivity. Indeed, these pronounced higher order effects were confirmed by supplementary simulations in which a free water component was added (supplementary Fig. S4). The supplementary simulations incorporating synthetic noise not only confirmed CTI's low precision for $b_{max}$ values lower than 1.5 ms/μm2 (for plausible diffusion scenarios in neural tissues, at least) but also showed that CTI's accuracy can also be affected by Rician noise biases (supplementary Fig. S3). Although Rician biases may not be relevant for our *ex vivo* results due the high SNR, these biases may explain the higher $K_{iso}$ estimates in inferior grey matter regions of the *in vivo* data which presented lower SNRs (Fig.6D, red arrows).

Given their observed b-value dependence, and similarly to all other metrics derived from the cumulant expansion, CTI measures cannot be considered to be completely accurate in comparison to their ground truth values. Nevertheless, our results show that CTI still provides a robust characterization of the isotropic and anisotropic kurtosis sources. Specifically, despite the imperfect accuracy, $K_{aniso}$ would still be sensitive to differences in microscopic anisotropy and clearly highlights regions of high anisotropy, while $K_{iso}$ would still be sensitive tensor magnitude variance across different compartments. Note that this observation is expected to be correct even for voxels containing partial volume effects (e.g., with free water) in which higher order effects are expected to be highly pronounced - according to our simulations, $K_{iso}$ in voxels containing free water (Fig. S4) is higher than the $K_{iso}$ estimates of voxels with no added free water (Fig. 7). Thus, CTI is able to robustly characterize the correct trends in the investigated system, also in terms of $K_{iso}$ estimates.

Regarding the $K_{intra}$ estimates, our simulations confirm that this metric is sensitive to restricted diffusion. Particularly, Fig. 7A4 shows that, for a given $b_{max}$ and a fixed degree of orientation dispersion, lower $K_{intra}$ estimates are still associated with bigger compartments. However, our simulations also revealed that the specific protocol used here for CTI is susceptible to higher order effects, and may introduce significant biases in $K_{intra}$ estimates (in some cases, higher than the ground truth by orders of magnitude). Since these biases depend on confounding effects such as degree of mesoscopic orientation dispersion (Fig. 7B4), higher order terms can compromise the specificity of $K_{intra}$ extracted using the current CTI protocol. Despite this issue, it is important to stress that $K_{intra}$ estimates reported in this study represent the first attempt of measuring intra-compartmental kurtosis separately from the anisotropic and isotropic kurtosis sources. We expect that our study will motivate further developments in CTI's framework to minimize confounding effects on $K_{intra}$ estimates (*vide infra*).

## 4.4. Future CTI vistas and clinical relevance

Recent studies based on isotropic diffusion encoding strategies (e.g. QTE) are showing that resolving different kurtosis sources can be clinically useful, because they can distinguish tissues with different underlying microstructural features such as different tumor types (Nilsson et al., 2020; Szczepankiewicz et al., 2016, 2015). Since the CTI approach provides a more general and less assumption-driven strategy to decouple these different kurtosis sources, it can potentially provide a more robust characterization of such tumours or of other disorders in which microscopic features are important. For instance, CTI-driven $K_{aniso}$ may more robustly detect the higher microscopic anisotropy of meningioma cells in comparison to the expected low anisotropy of glioblastoma cells. Previous studies have shown that diffusional kurtosis is a highly sensitive marker in stroke patients with stroke (Cheung et al., 2012; Rudrapatna et al., 2014). The different kurtosis sources afforded by CTI can likely provide important insights on, for example, the different microscopic/mesoscopic mechanisms underlying stroke, which are currently still under debate (Budde and Frank, 2010; Moseley et al., 1990b; van der Toorn et al., 1996). For basic research studies, resolving different kurtosis sources can be a valuable technique for decoupling microstructural features from mesoscopic confounding factors. For example, CTI can potentially be applied in studies of brain aging to decouple the effects of white and grey matter degeneration (likely detected by decreases of $K_{aniso}$) from the known increases of free water partial volume effects due to the gross morphological atrophy (likely detected by increases of $K_{iso}$) (Henriques, 2018; Metzler-Baddeley et al., 2012). CTI-driven $K_{iso}$ could also be sensitive to edema in neurodegenerative diseases, such as Parkinson diseases and Multiple Sclerosis, or in traumatic injury (Donkin and Vink, 2010; Gelfand et al., 2012; Lee et al., 2006). In addition to $K_{aniso}$ and $K_{iso}$ estimates, intra-compartmental kurtosis may be relevant in the future for investigating pathologies or processes in which microstructural modulations occur (e.g. axonal injury). Moreover, measuring intra-compartmental kurtosis via

CTI could be of interest for validation of other diffusion MRI methods that rely on assumptions such as vanishing intra-compartmental kurtosis or time-independent diffusion in tissues (Henriques et al., 2015; Jespersen et al., 2007; Novikov et al., 2019, 2018a; Szczepankiewicz et al., 2016, 2015; Westin et al., 2016; Zhang et al., 2012). We once more reiterate the importance of attenuating the higher order effects for resolving more accurate $K_{intra}$ values.

All these potential clinical and basic research applications motivate the future application of CTI to investigate porous systems at large and neural tissues in particular, as well as translation of CTI to clinical scanners. Although this study focused on CTI's proof-of-concept, our first *in vivo* contrasts of the rat brain show that consistent kurtosis source maps can be obtained from living animals (c.f. Fig. 6). As done for previous microscopic anisotropy measurements of DDE (Kerkelä et al., 2019; Yang et al., 2018) and multi-dimensional diffusion encoding (Sjölund et al., 2015), the clinical feasibility of CTI can be further promoted in future studies by refining its acquisition parameters and finding more optimal trends between precision, accuracy, and acquisition time. Although in this study, DDE data was processed with minimal pre-processing steps (only sub-pixel realigning was performed to correct data from motion and signals drifts), the CTI methodology can also be further improved by incorporating state-of-the-art denoising and artefact suppression algorithms. Even if future studies suggest that the CTI methodology is not yet compatible with clinical scanning acquisition times, the translation of the more general CTI approach to clinical scanners could still be fundamental to validate and calibrate the faster QTE acquisitions under different experimental conditions.

### 4.5. Methodological Limitations and further developments

As any other technique based on the cumulant expansion of diffusion-weighted signal decays (Chuhutin et al., 2017; Ianuş et al., 2018), CTI measures are biased by higher order terms.

Though their accuracy may be affected, CTI's $K_{aniso}$ and $K_{iso}$ may still be very useful if they provide a precise characterization of the different diffusional kurtosis sources, as mentioned above. By contrast, $K_{intra}$ estimates obtained using the current CTI protocol are more severely affected by higher order effects as mentioned repeatedly above and shown explicitly in (Fig. 7A4 and Fig.8A4), leading to potentially more confounds in its interpretation. Still, we expect that these higher order effects can be mitigated significantly. This can be achieved by, e.g.: 1) designing DDE b-value combinations that decrease the high order effects differences between the kurtosis tensor $W$ and the covariance tensor $C$ - since $K_{intra}$ information is captured by the subtraction of these tensors, minimizing the high order effects differences of these tensors can effectively suppress the biases in $K_{intra}$ estimates; 2) applying CTI after powder-averaging DDE signals, thereby effectively making them independent to tissue dispersion (Eriksson et al., 2015; Jespersen et al., 2013; Kaden et al., 2016; Lasič et al., 2014) - although this strategy may not eliminate the biases of higher order terms, it can remove the confounding dependence on orientation dispersion on all CTI metrics; 3) incorporating parameters in the CTI equation to fit and remove the impact of higher order terms, such as previously done for microscopic anisotropy estimates from DDE powder-averaged signals (Ianuş et al., 2018). Note that future studies should carefully assess the robustness of such bias minimization strategies since they might also affect CTI's precision. For instance, previous studies DKI studies, showed that incorporating higher order terms in the cumulant expansion equation significantly compromises the precision of the lower order terms (Chuhutin et al., 2017; Kiselev, 2017). These mitigation strategies are currently being implemented and will be reported in due course.

# 5. Conclusion

We provide a first general framework for measuring the diffusion correlation tensor directly from double diffusion encoded signals and without a-priori assumptions. The ensuing Correlation Tensor Imaging (CTI) approach was here theoretically derived, explored via simulations and applied to characterize the different sources of diffusional kurtosis in rodent brains both *ex vivo* and *in vivo*. Our theory shows that CTI resolves kurtosis sources, and can provide quantitative indices related to the anisotropic and isotropic diffusion variances ($K_{aniso}$ and $K_{iso}$) without relying on the Gaussian diffusion assumption; in addition, CTI offers an index sensitive to intra-compartmental kurtosis ($K_{intra}$) arising from restricted diffusion. Although future CTI protocol optimization is still required for mitigating higher order term effects on $K_{intra}$ estimates, our results suggest that separating kurtosis sources is a promising vista for imparting specificity on kurtosis measurements, without relying on specific microstructural assumption or constraints. All these features augur well for future development and implementation of CTI measures for basic research and biomedical applications.

# Acknowledgments

This study was funded by the European Research Council (ERC) (agreement No. 679058). The authors acknowledge the vivarium of the Champalimaud Centre for the Unknow, a facility of CONGENTO which is a research infrastructure co-financed by Lisboa Regional Operational Programme (Lisboa 2020), under the PORTUGAL 2020 Partnership Agreement through the European Regional Development Fund (ERDF) and Fundacao para a cienecia e tecnologia (Portugal), project LISBOA-01-0145-FEDER-022170. The authors also want to thank Prof Dr Valerij G. Kiselev (Freiburg University) and Mr. Leevi Kerkela (UCL) for insightful discussions and suggestions, Ms. Teresa Serradas Duarte and Dr. Daniel Nunes for assistance in the preparation of the *ex vivo* mouse brain specimens, and Andrada Ianus for sharing her previous DDE sequence implementations.

# Figures

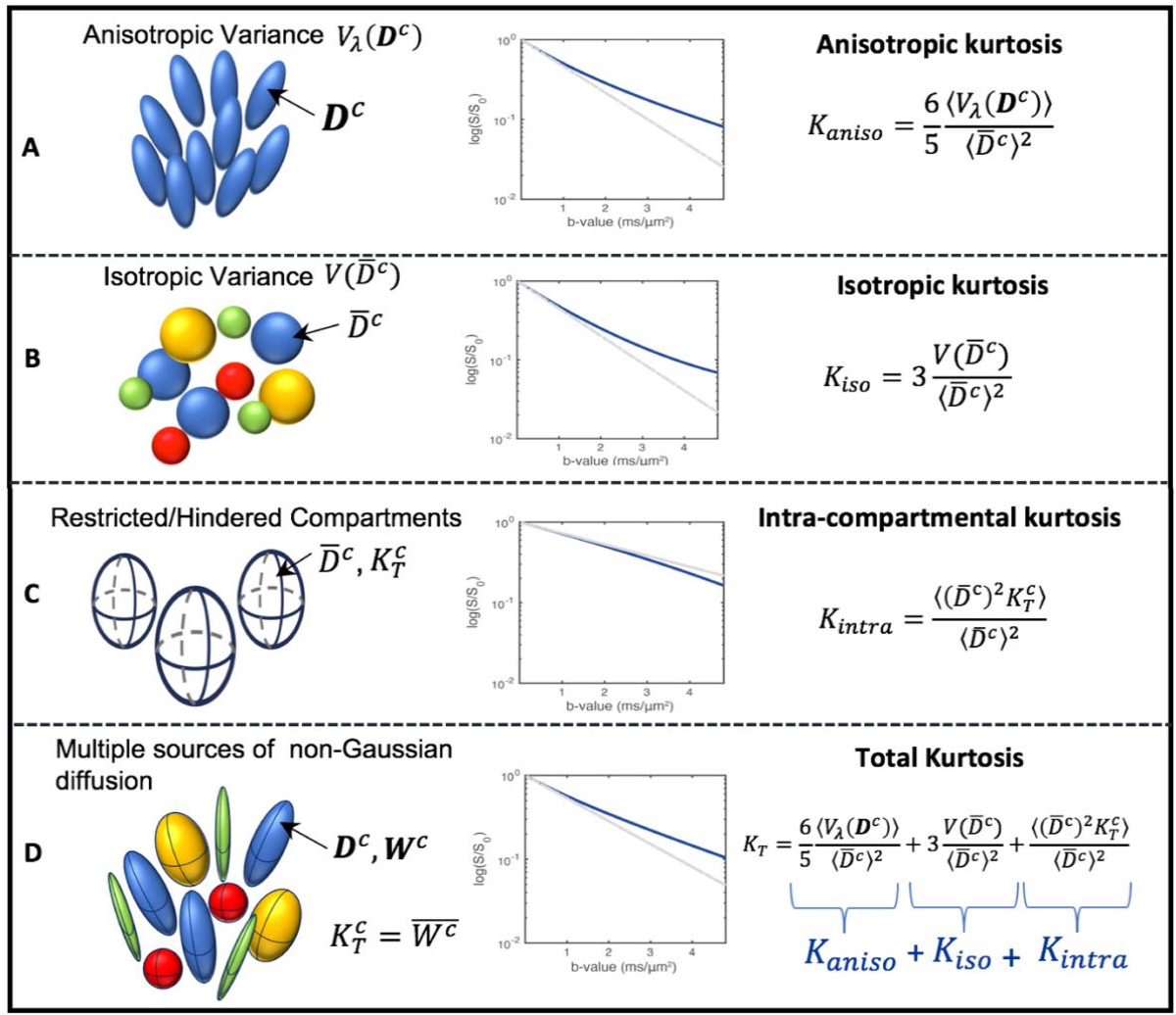

**Figure 1 – Illustration of inter-compartmental and intra-compartmental kurtosis sources**. **A)** Kurtosis can emerge in the dMRI signal decay if replicas of a single diffusion tensor ($D^c$) are mesoscopically dispersed along multiple orientations. In this scenario, kurtosis emerges from the variance across the different eigenvalues of the individual diffusion tensor $D^c$, and is hence termed here "anisotropic kurtosis". **B)** Kurtosis can also arise from polydisperse tensor traces, without orientation dispersion. In this scenario, kurtosis can be fully determined by the diffusion variance across the different mean diffusivities $\bar{D}^c$; hence, it is termed isotropic kurtosis. **C)** Finally, kurtosis can be a result of non-Gaussian, restricted diffusion within individual compartments with reflecting barriers. This kurtosis source is termed here intra-compartmental kurtosis. Given that (A) and (B) represent ensemble properties (n.b., if the gaussian diffusion tensors comprising the ensemble were isolated, each individual component would not exhibit any kurtosis), they can be considered as "inter-compartmental" kurtosis sources; (C) emerges even in a single compartment. **D)** In realistic tissues, diffusional kurtosis can be expected to emerge from a combination of all the different sources of inter- and intra-compartmental kurtosis - the total kurtosis is the sum of the three above kurtosis sources.

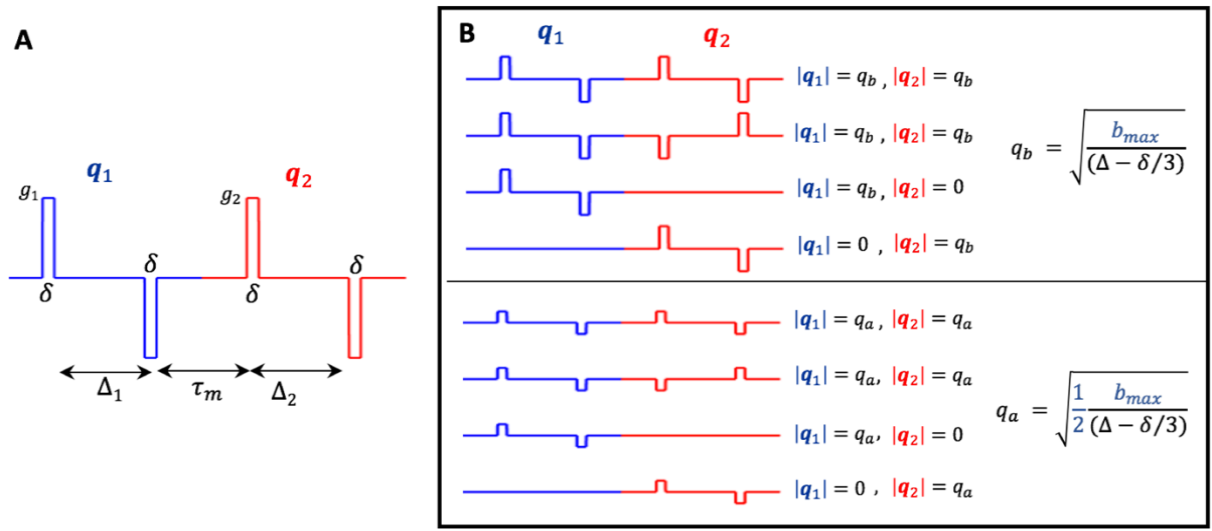

**Figure 2 – Acquisition requirements for a minimal protocol designed for CTI: A)** Parameters and waveform of a standard DDE pulse sequence ($\Delta_1$ and $\Delta_2$ are the diffusion gradient separation times, $\delta$ is the diffusion gradient pulse duration, and $\tau_m$ is the mixing time between the two diffusion encoding modules marked by the blue and red lines); **B)** The eight gradient intensity combinations used for the minimal protocol of CTI. These gradient intensity combinations are acquired for 117 directions pairs for $q_1$-$q_2$ directions (Jespersen's 5 design + 45 parallel DDE experiments) and can be acquired for different $b_{max}$ values.

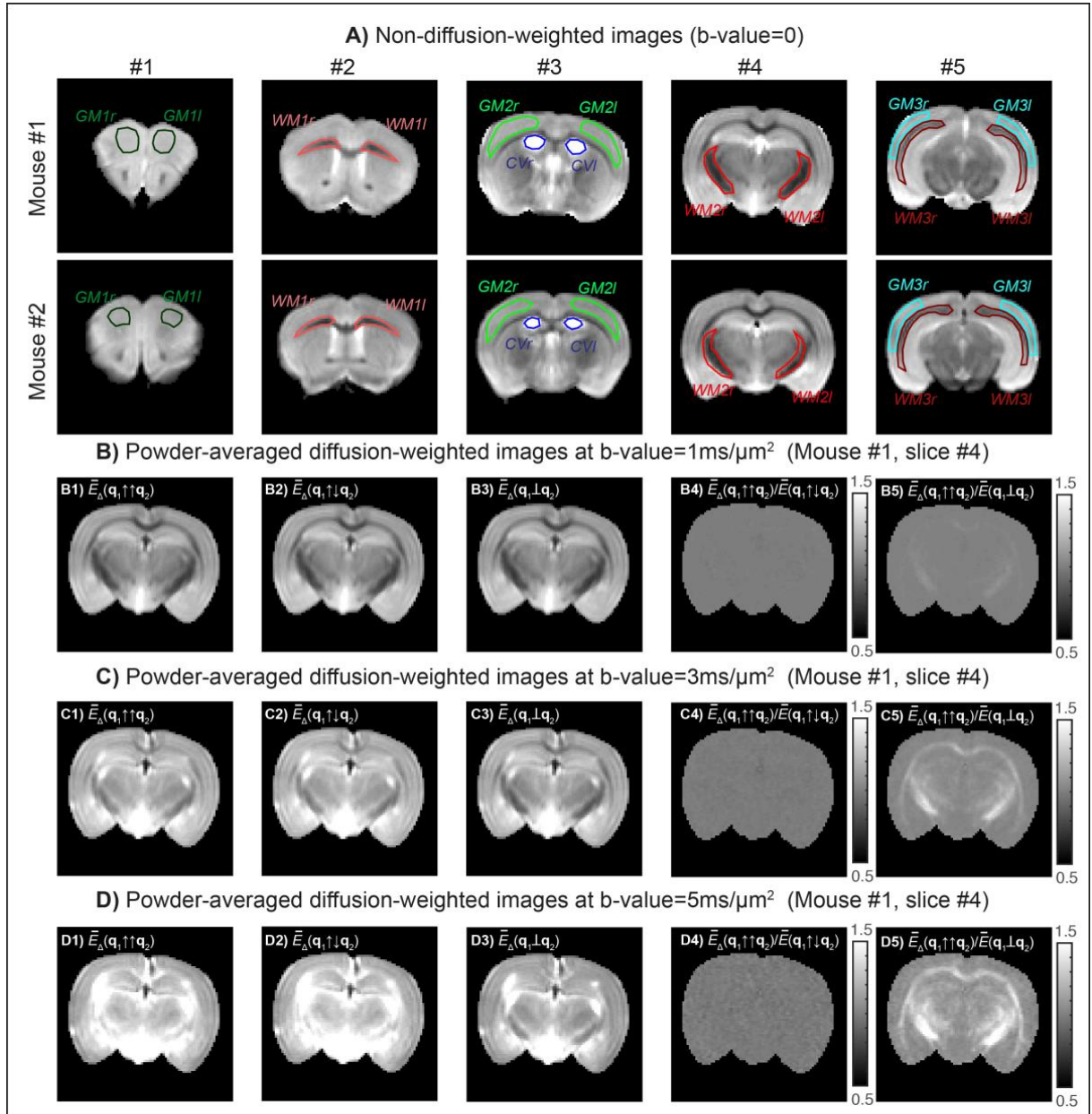

**Figure 3 – Raw double diffusion encoding (DDE) data: A)** Non-diffusion weighted images of the five coronal slices acquired from both mouse brains; regions of interest (ROIs) were manually defined. White matter ROIs were drawn in corpus callosum white matter (*WM1r* and *WM1l*), internal capsule white matter (*WM2r* and *WM2l*), and external capsule white matter (*WM3r* and *WM3l*), while GM ROIs comprised grey matter of motor cortex (*GM1r* and *GM1l*), somatosensory cortex (*GM2r* and *GM2l*), visual cortex, and auditory cortex (*GM3r* and *WM3l*). For reference, ROIs were also drawn in the cerebral ventricles (large contribution of free water diffusion, *CVr* and *CVl*); **B)** Powder average DDE data decays at a total b-value = 1 ms/μm² for slice #4 of mouse specimen #1 and for the following experiments: B1) parallel DDE experiments; B2) anti-parallel DDE experiments; B3) perpendicular DDE experiments; B4) ratio between parallel and anti-parallel experiments; and B5) ratio between parallel and perpendicular experiments; **C)** Powder averaged DDE signal decays at a total b-value = 3 ms/μm² for slice #4 of mouse specimen #1 and for: C1) parallel DDE experiments; C2) anti-parallel DDE experiments; C3) perpendicular DDE experiments; C4) ratio between parallel and anti-parallel experiments; and C5) ratio between parallel and perpendicular experiments; **D)** Powder averaged DDE signal decays at total b-value = 5 ms/μm² for slice #4 of mouse specimen #1 and for: D1) parallel DDE experiments; D2) anti-parallel DDE experiments; D3) perpendicular DDE experiments; D4) ratio between parallel and anti-parallel experiments; and D5) ratio between parallel and perpendicular experiments.

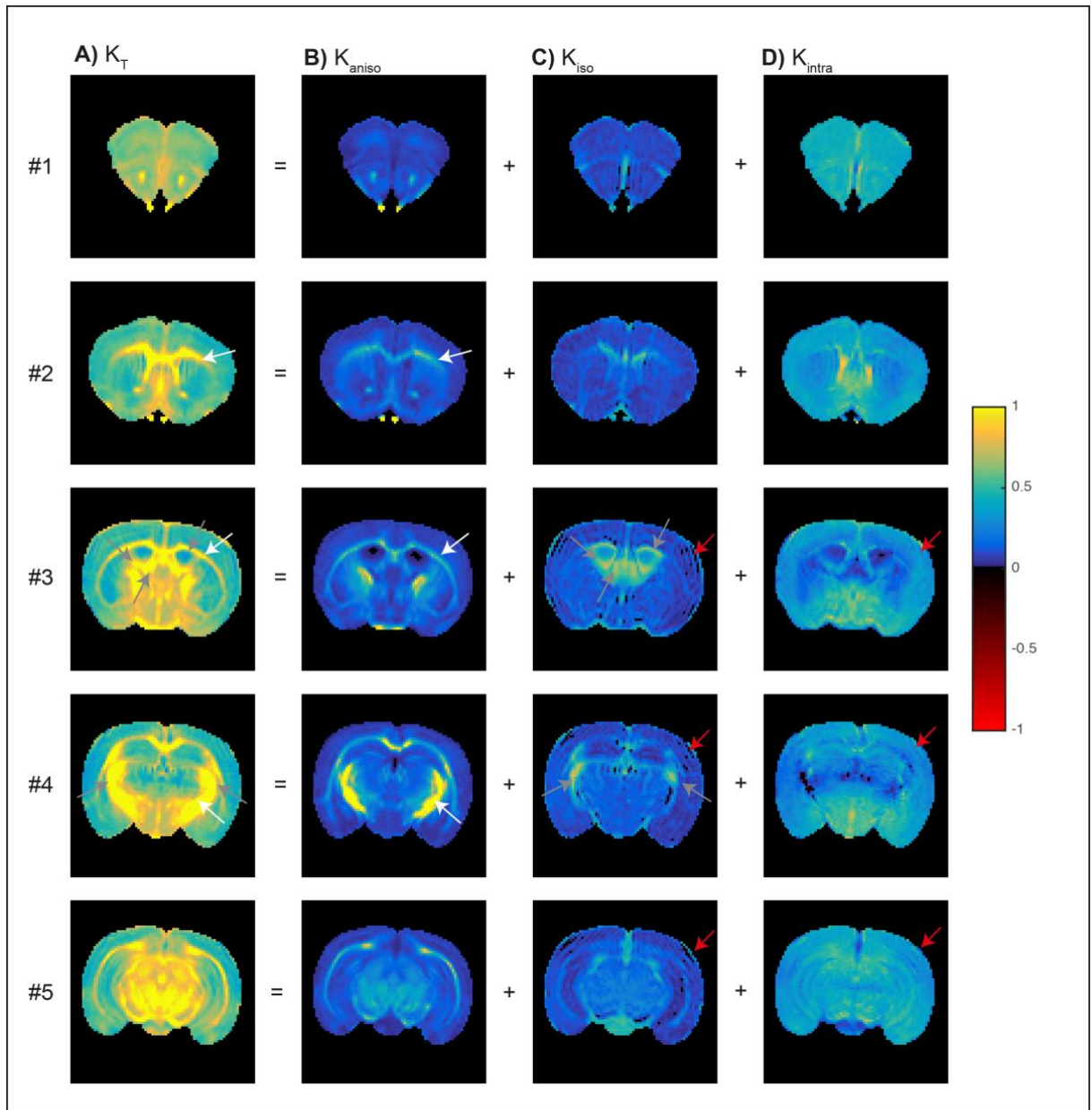

**Figure 4 – CTI kurtosis measures for all five slices extracted from all data acquired in mouse brain specimen #1. A)** total kurtosis of powder-average signals; **B)** anisotropic kurtosis; **C)** isotropic kurtosis; and **D)** intra-compartmental kurtosis. White arrows highlight white matter regions with high anisotropy; grey arrows point to regions contaminated by free water partial volume effects; and red arrows point towards regions corrupted by Gibbs ringing artefacts.

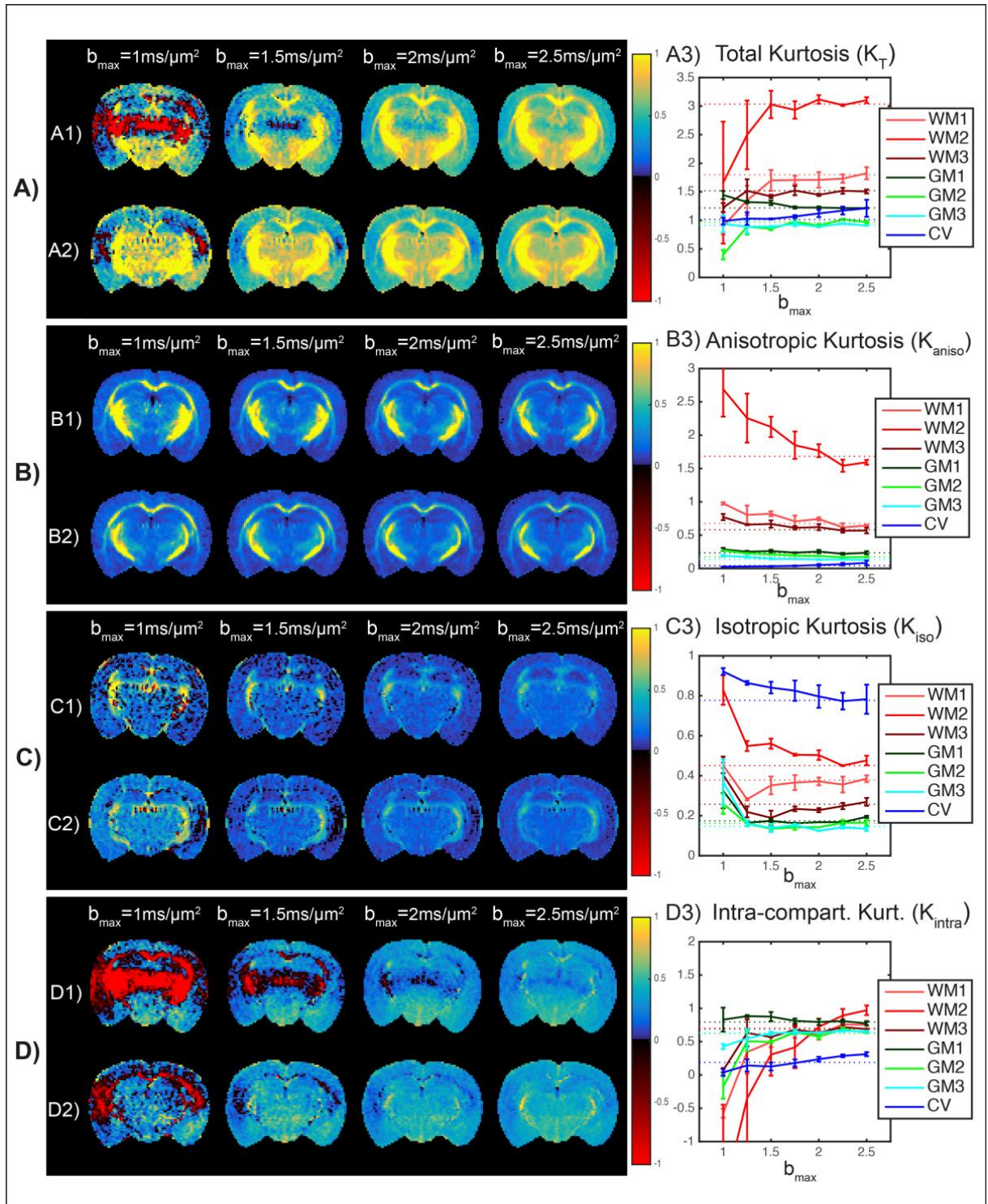

**Figure 5 – CTI kurtosis measures extracted from minimal protocols for different $b_{max}$ values. A)** total kurtosis; **B)** anisotropic kurtosis; **C)** isotropic kurtosis; and **D)** intra-compartmental kurtosis. Parametric maps in the left of each panel are plotted for sub-protocols with $b_{max}$ = 1, 1.5, 2, 2.5 ms/μm2 and for the mouse brain specimen #1 (A1, B1, C1, and D1) and mouse brain specimen #2 (A2, B2, C2, and D2); while the kurtosis mean and standard deviation across animals for seven ROIs are shown in panels A3, B3, C3, and D3. For a reference, the mean kurtosis estimates of the extensive sampled b-value protocol are plotted by the dotted lines in panels A3, B3, C3, and D3.

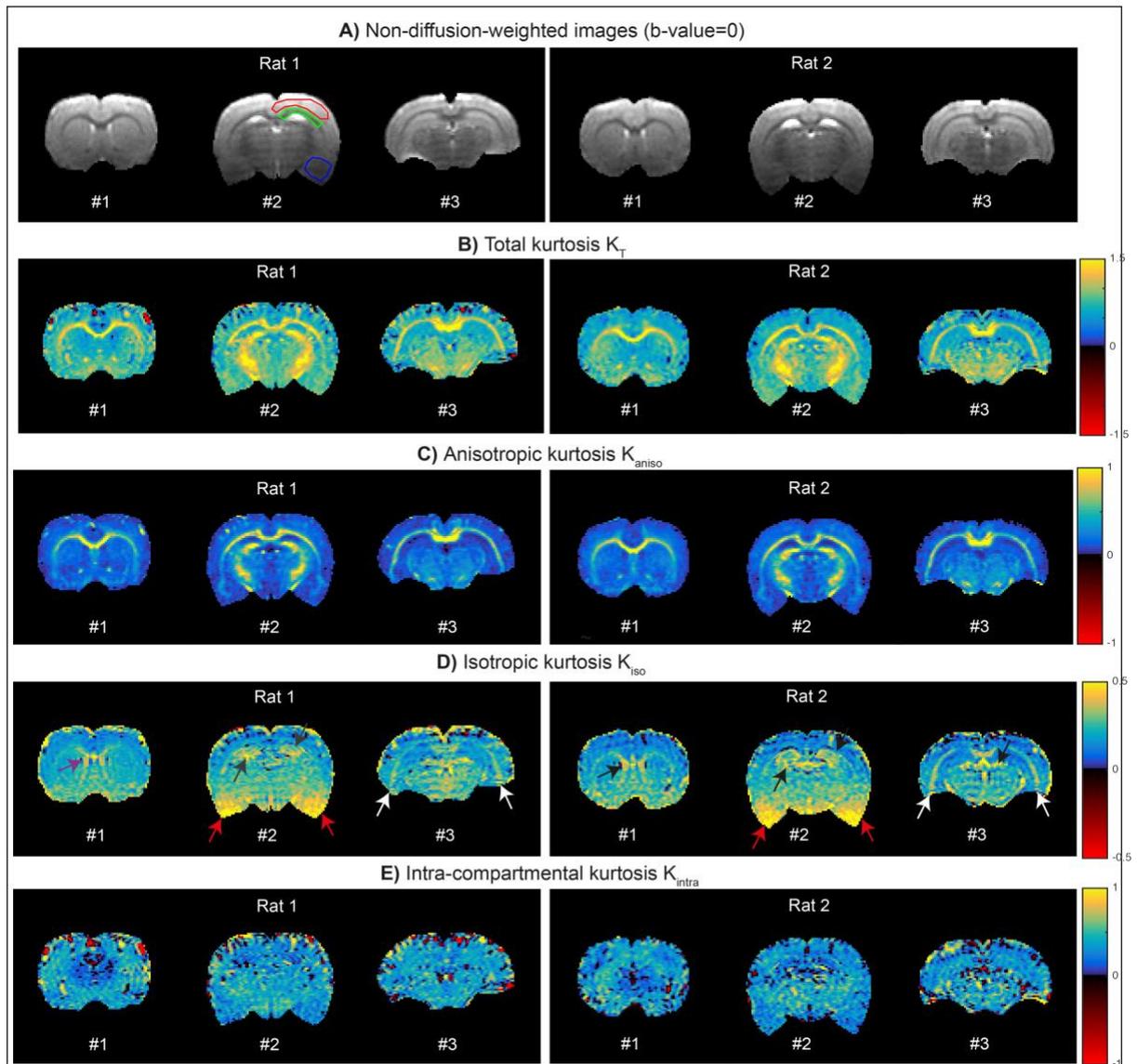

**Figure 6 – CTI kurtosis measures extracted from *in vivo* rat brain data. A)** Non-diffusion weighted images for all three acquired slices of both rats (white and grey matter ROIs are manually defined in slice #2 of rat brain #1 for SNR estimation); **B)** maps of the total kurtosis for all three acquired slices in both rats; **C)** maps of the anisotropic kurtosis for all three acquired slices in both rats; **D)** maps of the isotropic kurtosis for all three acquired slices of both rats (magenta arrows point to areas where partial volume effects between tissue and free water of cerebral ventricles is high; white arrows point to white matter); and **E)** maps of the intra-compartmental kurtosis for all three acquired slices in both rats.

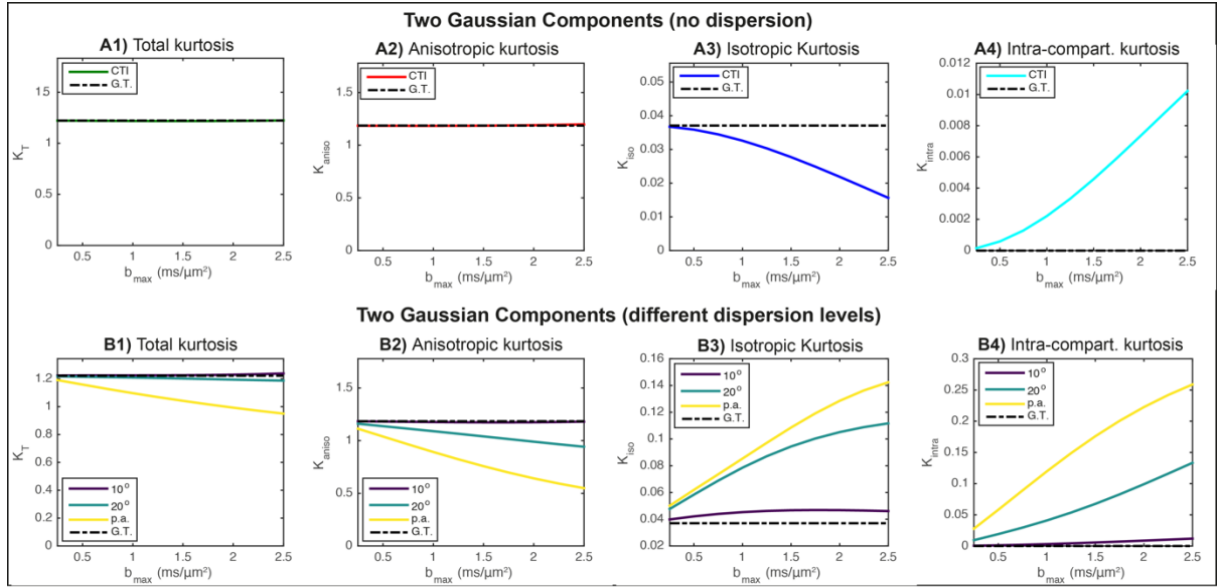

**Figure 7 – CTI kurtosis measures for synthetic signals of environments containing two types of Gaussian components. A)** Simulations performed based on two aligned Gaussian components (axial and radial diffusivities for the first component are 2 and 0 μm²/ms, respectively, while the axial and radial diffusivities for the second compartment are 1.5 and 0.5 μm²/ms, respectively) - total kurtosis, anisotropic kurtosis, isotropic kurtosis, and intra-compartmental kurtosis estimates are plotted as a function of $b_{max}$ from panels A1 to A4, respectively. **B)** Simulations performed based on replicas of the two Gaussian components dispersing at different degrees (dispersion angles of 10° and 20° are plotted with purple and green lines, while completely powder-averaged (p.a.) replicas are plotted with the yellow line) - total kurtosis, anisotropic kurtosis, isotropic kurtosis, and intra-compartmental kurtosis estimates are plotted as a function of $b_{max}$ in panels B1 to B4, respectively. Ground truth values are marked by the black dashed lines. Note that individual kurtosis estimates in these curves were obtained using 8 $q_1$ and $q_2$ gradient intensity combinations which are fully defined given a single $b_{max}$ value according to the minimal CTI protocol.

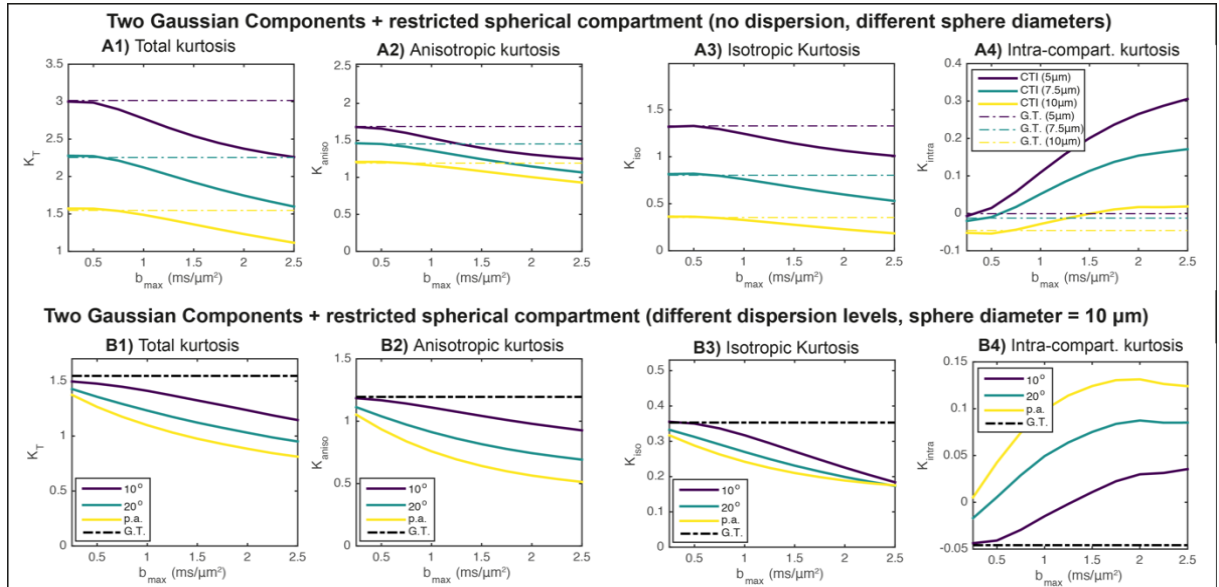

**Figure 8 – CTI kurtosis metrics for synthetic signals of environments containing two types of Gaussian components and a spherical compartment in which restricted diffusion occurs. A)** Simulations based on two coherently aligned Gaussian components and a restricted spherical compartment for different diameters (spherical diameters of 5, 7.5, and 10 μm are plotted in purple, green and yellow lines, respectively) - total kurtosis, anisotropic kurtosis, isotropic kurtosis, and intra-compartmental kurtosis estimates are plotted as a function of $b_{max}$ from panels A1 to A4. **B)** Simulations performed based on replicas of two Gaussian components with different levels of dispersion and a restricting spherical compartment with diameter of 10 μm (dispersion angles of 10º and 20º are plotted with purple and green lines, while completely powder-averaged (p.a.) replicas are plotted with the yellow line)- total kurtosis, anisotropic kurtosis, isotropic kurtosis, and intra-compartmental kurtosis estimates are plotted as a function of $b_{max}$ in panels B1 to B4, respectively. Ground truth values are marked by dashed lines. For these simulations, the different underlying components were set to have similar signal contributions. Note that individual kurtosis estimates in these curves are obtained using 8 $q_1$ and $q_2$ gradient intensity combinations which are fully defined given a single $b_{max}$ value according to the minimal CTI protocol.

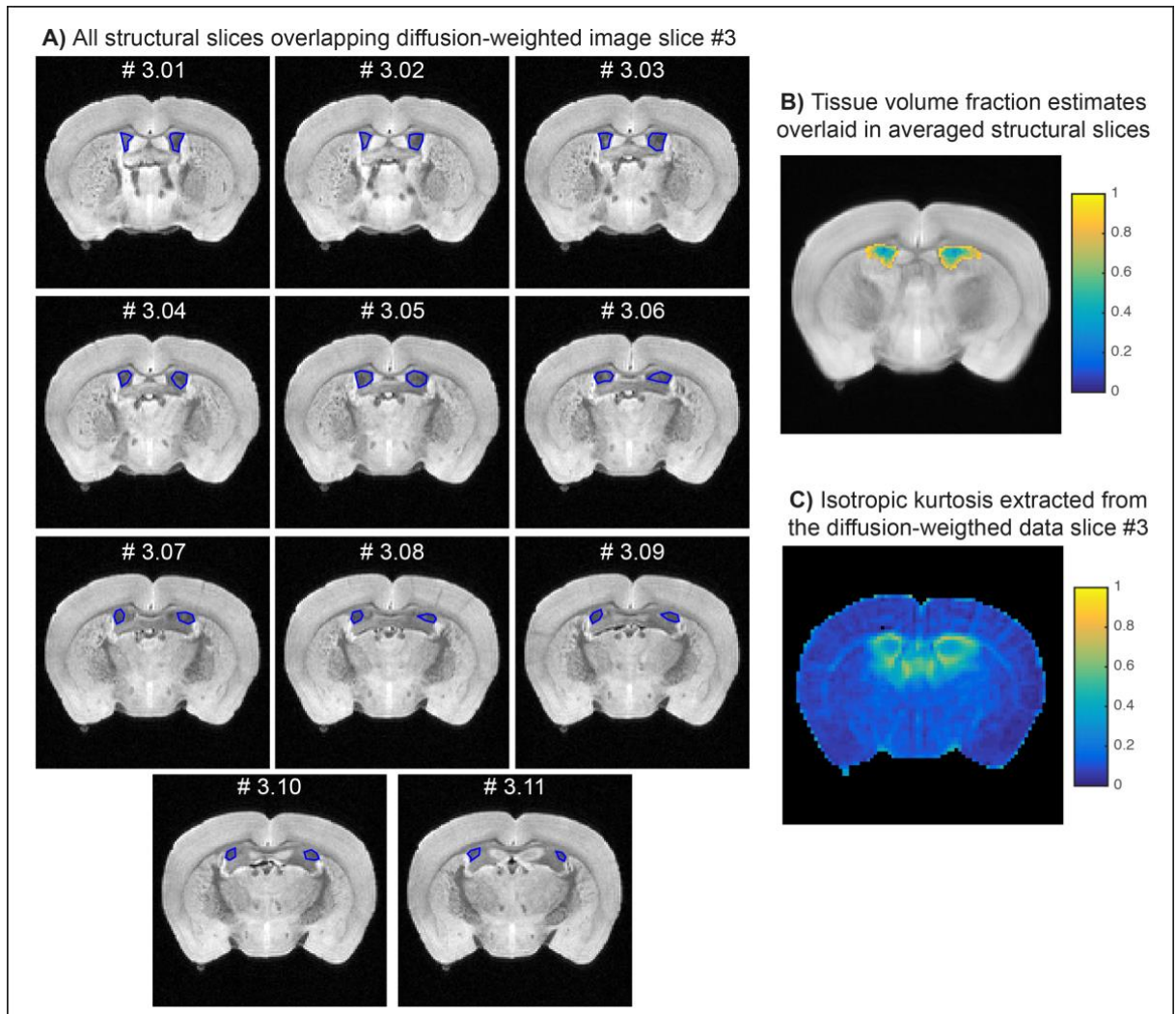

**Supplementary Figure S1 – Partial volume effects between tissue and cerebral ventricle as estimated from high-resolution data. A)** All high-resolution structural images (slice thickness = 79.4 μm) overlapping with the thicker diffusion-weighted image slice #3 (with thickness = 900 μm) – in all 11 structural images the cerebral ventricles (CV) were manually outlined (blue outlines on each panel). **B)** Tissue volume fraction map overlaid in averaged high-resolution structural image – tissue volume fraction is estimated by first projecting the 11 CV ROIs to the averaged structural imaging and then dividing the number of non-overlaying ROIs at each voxel position by the total number of CV ROIs (i.e. 11) – these volume fraction estimates are only performed on voxels overlaid by at least one ROI. **C)** For comparison, isotropic kurtosis estimated from diffusion-weighted data of slice #3 is displayed. The volume fraction profile in panel B is consistent with the profiles of the high isotropic kurtosis in panel C. This suggests that high isotropic kurtosis values close to the cerebral ventricles are a consequence of the partial volume effects between tissue and free water tissue. This figure was produced for mouse specimen #2.

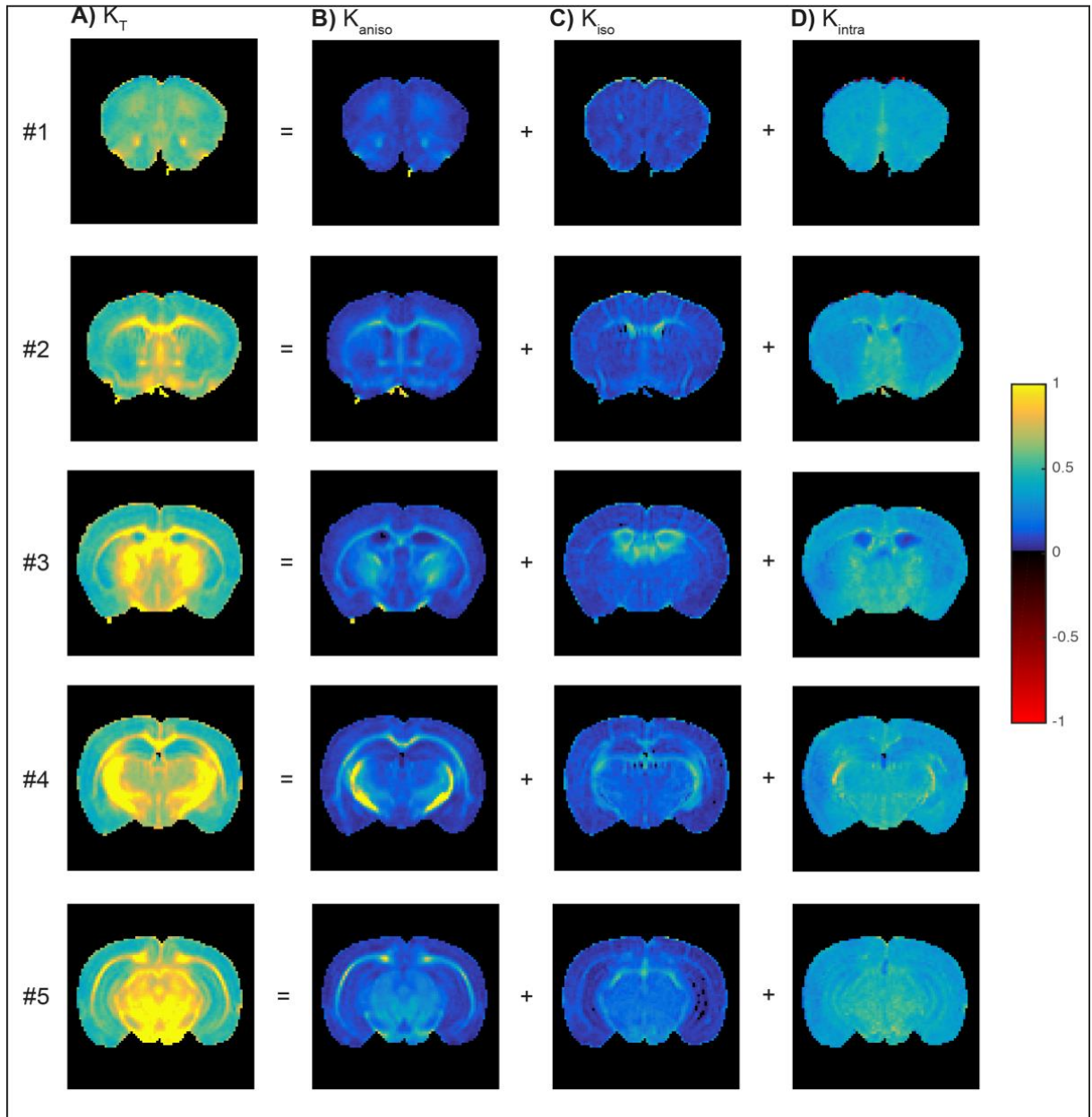

**Supplementary Figure S2 – CTI kurtosis measures for all five slices extracted from all data acquired in mouse brain specimen #2. A)** total kurtosis of powder-average signals; **B)** anisotropic kurtosis; **C)** isotropic kurtosis; and **D)** intra-compartmental kurtosis.

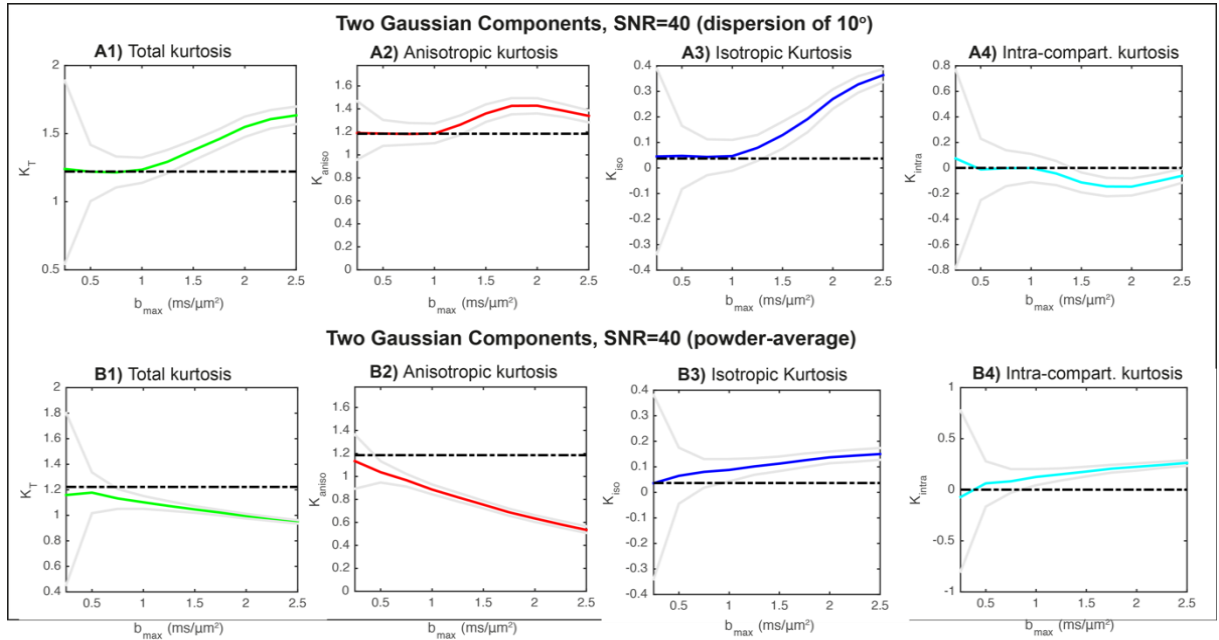

**Supplementary Figure S3 - CTI kurtosis measures for simulated signals corrupted with synthetic Rician distributed noise (SNR=40) for two tissue dispersion levels: A)** Simulations of two types of Gaussian components dispersing at a dispersion angles of $10°$- the median, 25th and 75th percentile values for total kurtosis, anisotropic kurtosis, isotropic kurtosis, and intra-compartmental kurtosis are plotted as a function of $b_{max}$ in panels A1 to A4, respectively - solid colored lines represents the median values, the 25th and 75th percentiles are plotted by the grey lines, and the ground truth values are marked by the black dashed lines. **B)** Simulations of two types of randomly oriented Gaussian components (i.e. powder-averaged compartments) – the median values of total kurtosis, anisotropic kurtosis, isotropic kurtosis, and intra-compartmental kurtosis estimates are plotted as a function of $b_{max}$ in panels B1 to B4, respectively. Note that: 1) the SNR of simulations was set according to the estimates obtained from the superior ROIs of the *in vivo* rat data; 2) Median and percentile values are computed for 10000 noise initializations; 3) individual kurtosis estimates for noise iteration are obtained using 8 $q_1$ and $q_2$ gradient intensity combinations which are fully defined given a single $b_{max}$ value according to the minimal CTI protocol.

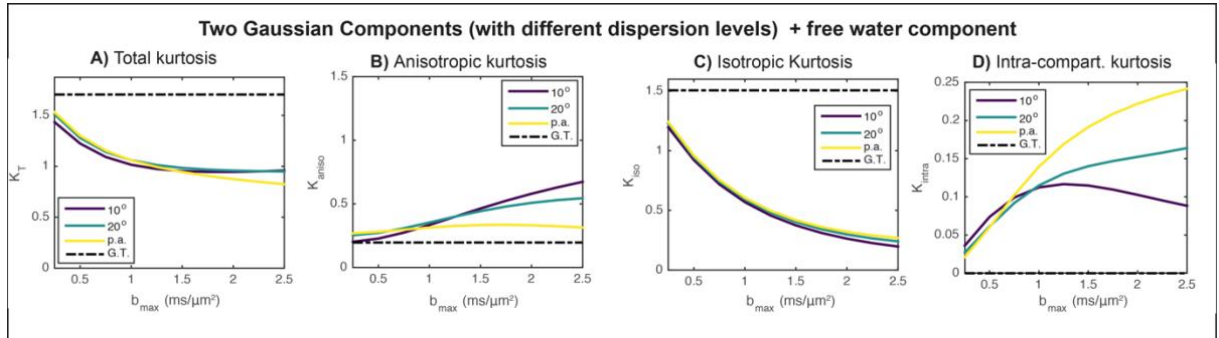

**Supplementary Figure S4 – CTI kurtosis measures for synthetic signals containing a free water component:** **A)** total kurtosis estimates; **B)** anisotropic kurtosis estimates; **C)** isotropic kurtosis estimates; and **D)** intra-compartmental kurtosis estimates. These simulations were performed based on replicas of two Gaussian components (components 1 and 2) dispersing at different degrees in addition to a free water Gaussian component. For these simulations, the different compartments were set to have similar signal contributions (i.e. volume fractions of 1/3), axial and radial diffusivities for the first component are 2 and 0 μm²/ms, axial and radial diffusivities for the second compartment are 1.5 and 0.5 μm²/ms, while the isotropic diffusivity for compartment 3 was set to 3 μm²/ms. In each panel, results for the dispersion angles of 10° and 20° are plotted in purple and green lines, while completely powder-averaged (p.a.) replicas are plotted in the yellow line. Ground truth values are marked in black dashed lines. Note that individual kurtosis estimates on these curves are obtained using 8 $q_1$ and $q_2$ gradient intensity combinations which are fully defined given a single $b_{max}$ value according to the minimal CTI protocol.